\documentclass[jcappub,11pt]{article}
\pdfoutput=1 
\usepackage{jcappub} 
\usepackage[T1]{fontenc} 
\usepackage{url}
\usepackage{xcolor}
\usepackage{graphicx}
\usepackage{multirow}
\usepackage{mathrsfs}
\usepackage[caption=false]{subfig}

\usepackage{tikz,xcolor,hyperref}

\definecolor{lime}{HTML}{A6CE39}
\DeclareRobustCommand{\orcidicon}{%
	\begin{tikzpicture}
	\draw[lime, fill=lime] (0,0) 
	circle [radius=0.16] 
	node[white] {{\fontfamily{qag}\selectfont \tiny ID}};
	\draw[white, fill=white] (-0.0625,0.095) 
	circle [radius=0.007];
	\end{tikzpicture}
	\hspace{-5mm}
}

\foreach \x in {A, ..., Z}{%
	\expandafter\xdef\csname orcid\x\endcsname{\noexpand\href{https://orcid.org/\csname orcidauthor\x\endcsname}{\noexpand\orcidicon}}
}

\title{\boldmath Cosmological constraints from the Minkowski functionals of the BOSS CMASS galaxy sample}

\author[a,b]{Wei Liu\orcidA{},}
\author[c]{Enrique Paillas\orcidB{},}
\author[d,e,f]{Carolina Cuesta-Lazaro\orcidC{},}
\author[g,h]{Georgios Valogiannis\orcidD{},}
\author[a,b,1]{and Wenjuan Fang\orcidE{} \note{Corresponding author.}}

\affiliation[a]{Department of Astronomy, University of Science and Technology of China, Hefei, Anhui, 230026, P.R.China}
\affiliation[b]{School of Astronomy and Space Sciences, University of Science and Technology of China, Hefei, Anhui, 230026, P.R.China}
\affiliation[c]{Department of Astronomy/Steward Observatory, University of Arizona, 933 North Cherry Avenue, Tucson, AZ 85721-0065, USA}
\affiliation[d]{The NSF AI Institute for Artificial Intelligence and Fundamental Interactions, Cambridge, MA 02139, USA}
\affiliation[e]{Department of Physics, Massachusetts Institute of Technology, Cambridge, MA 02139, USA}
\affiliation[f]{Center for Astrophysics — Harvard \& Smithsonian, 60 Garden Street, MS-16, Cambridge, MA 02138, USA}
\affiliation[g]{Department of Astronomy \& Astrophysics, University of Chicago, Chicago, IL, 60637, USA}
\affiliation[h]{Kavli Institute for Cosmological Physics, Chicago, IL, 60637, USA}

\emailAdd{lw980228@mail.ustc.edu.cn}
\emailAdd{epaillas@arizona.edu}
\emailAdd{carolina.cuesta-lazaro@cfa.harvard.edu}
\emailAdd{gvalogiannis@uchicago.edu}
\emailAdd{wjfang@ustc.edu.cn}

\abstract{For the first time, we develop a simulation-based model for the Minkowski functionals (MFs) of large-scale structure, which allows us to extract the full information available from the MFs (including both the Gaussian and non-Gaussian part), and apply it to the BOSS DR12 CMASS galaxy sample. Our model is based on high-fidelity mock galaxy catalogs constructed from the \textsc{Abacus}\textsc{Summit} simulations using the halo occupation distribution (HOD) framework, which include the redshift-space distortions and Alcock-Paczynski distortions, incorporate survey realism, including survey geometry and veto masks, and account for angular plus radial selection effects. The cosmological and HOD parameter dependence of the MFs is captured with a neural network emulator trained from the galaxy mocks with various cosmological and HOD parameters. To benchmark the constraining power of the MFs, we also train an emulator for the galaxy 2-point correlation function (2PCF) using the same pipeline. Having validated our approach through successful parameter recovery tests on both internal and external mocks, including non-HOD forward models of the halo-galaxy connection, we apply our forward model to analyze the CMASS data in the redshift range $0.45<z<0.58$. We find the MFs provide stronger constraints on the cosmological parameters than the 2PCF. The combination of the two gives $\omega_{\rm cdm}=0.1172^{+0.0020}_{-0.0023}$, $\sigma_8=0.783\pm 0.026$, and $n_s=0.966^{+0.019}_{-0.015}$, which are tighter by a factor of 2.0, 1.9, and 1.6 than the 2PCF alone. The derived constraint $f\sigma_8=0.453 \pm 0.016$ is also improved by a factor of 1.9, compared to the 2PCF, and agrees well with Planck 2018 predictions and other results from a series of studies in the literature. This work provides a new methodology for the application of the MFs to galaxy surveys and demonstrates that non-Gaussian information embedded in the MFs can be exploited to obtain strong constraints on cosmological parameters.
}

\begin{document}
\maketitle
\flushbottom

\section{Introduction}
The topology of large-scale structure (LSS) traced by the galaxies has always been an interesting topic in cosmology, besides the clustering of galaxies. After the finding of the sponge-like topology of cosmic LSS, which was revealed by the iso-density contours and the genus statistic of the CfA catalog \cite{1986ApJ...306..341G}, a series of follow-up works \cite{1987ApJ...319....1G,1987ApJ...321....2W,1988PASP..100.1307G,1988ApJ...328...50M,1989ApJ...340..625G} directly compared the genus measured for the CfA redshift survey to that predicted by the oldfashioned cosmological and galaxy formation models then. It was soon realized by the cosmology community that the genus, or equivalently, the Euler characteristic, is one of the Minkowski functionals \cite{Minkowski1903,1994A&A...288..697M}, which provide a unique and complete way to characterize the morphology of cosmic structure. According to Hadwidger's theorem \cite{Hadwiger_1957}, for a pattern in n-dimensional space, its morphological properties (defined as those satisfying motional-invariance and additivity) can be fully described by (n+1) Minkowski functionals (MFs) \footnote{The more rigorous and mathematical description of Hadwiger's theorem is given in \cite{1994A&A...288..697M} as: any additive, motion invariant and conditionally continuous functional $\mathscr{F}$ on a body $A$ in $d$ dimension is a linear combination of the $d+1$ Minkowski functional $\mathscr{F}(A)=\sum_{\i=0}^d c_i V_i(A)$, with real coefficients $c_i$ independent of $A$.}. In 3D, the 4 MFs are, respectively, the pattern's volume, surface area, integrated mean curvature, and Euler characteristic (or genus).

After their introduction into cosmology by \cite{1994A&A...288..697M,1997ApJ...482L...1S}, the MFs (the works focus on the genus statistics are also included) are extensively applied to observed galaxy samples including the IRAS Point Source Catalogue Redshift Survey \cite{1998MNRAS.297..777C}, the CfA2 redshift survey \cite{2000MNRAS.312..638S}, the Two-Degree Field Galaxy Redshift Survey \cite{2002ApJ...570...44H}, the Sloan Digital Sky Survey (SDSS) Early Data Release galaxy sample \cite{2002PASJ...54..707H}, the Large-scale Structure Sample 12 \cite{Blanton_2003} of the SDSS galaxy redshift data \cite{10.1093/pasj/55.5.911}, the New York University Value-Added Galaxy Catalog \cite{2005AJ....129.2562B} constructed from the SDSS \cite{2005ApJ...633...11P,2008ApJ...675...16G}, the Seventh Data Release of the SDSS \cite{2010ApJS..190..181C,2010ApJ...722..812Z}, and the CMASS Data Release 10 (DR10) sample \cite{Eisenstein_2011,2014ApJS..211...17A} of the SDSS-III Baryon Oscillation Spectroscopic Survey (BOSS) \cite{2014ApJ...796...86P}. These earlier applications of the MFs mainly focus on the effects of a variety of observational systematics, the dependence on the Gaussian smoothing scale and on the intrinsic properties of galaxies like luminosity and morphology, and the comparison of the MFs measured from the observed galaxies and the state-of-the-art cosmological N-body simulations at that time. They also discussed the difference in the MFs between the measurement for the observed data and the so-called Tomita’s formula for MFs of Gaussian random fields \cite{Tomita_1990}, where the amplitudes of the MFs are the only free parameters and can be calculated from the power spectrum of the density field.

Other interesting topics and methodologies have also been explored with the application of the MFs to observed galaxies. Based on the Germ–Grain method \cite{2014MNRAS.443..241W} (in contrast to the more frequently used isodensity contour method) for the measurement of the Minkowski Functionals, \cite{2017MNRAS.467.3361W} first probed the higher-order clustering of the galaxies using the analytically known connection of the MFs to the integrals over higher-order correlation functions. With the same methodology, \cite{2019MNRAS.485.1708S} studied the evolution of higher-order correlations extracted from the Minkowski functionals.

Constraints on cosmological parameters can be derived from the observed data with the MFs. The idea of using the redshift dependence of the genus amplitude as a standard ruler was originally proposed in \cite{2010ApJ...715L.185P,10.1111/j.1365-2966.2010.18015.x}. This method was applied to the WiggleZ survey \cite{10.1111/j.1365-2966.2011.19077.x} first by \cite{10.1093/mnras/stt2062} and then to the SDSS-III BOSS catalog by \cite{2021ApJ...907...75A}. On the other hand, the genus amplitude itself, together with the amplitude of other orders of the MFs, provides a measure of the shape of the linear matter power spectrum and can thus constrain cosmological parameters \cite{Appleby_2020,2022ApJ...928..108A}. However, large smoothing scales were used in these works to reduce non-Gaussian and nonlinear effects, which are hard to model or correct. Therefore, only the Gaussian information was exploited to obtain cosmological constraints. 

 Current and upcoming galaxy surveys such as DESI \footnote{\href{http://www.desi.lbl.gov}{http://www.desi.lbl.gov}}, PFS \footnote{\href{http://pfs.ipmu.jp}{http://pfs.ipmu.jp}}, the Nancy Grace Roman Space Telescope \footnote{\href{https://science.nasa.gov/mission/roman-space-telescope}{https://science.nasa.gov/mission/roman-space-telescope}}, Euclid \footnote{\href{http://sci.esa.int/euclid}{http://sci.esa.int/euclid}} and CSST \cite{CSST,2019ApJ...883..203G}\footnote{\href{http://nao.cas.cn/csst}{http://nao.cas.cn/csst}}, will provide high-precision measurements of the 3D clustering of galaxies. To fully extract and exploit the wealth of valuable information provided by these cosmological observations, statistical tools beyond the standard 2-point correlation function and power spectrum are needed to probe non-Gaussian information induced by nonlinear processes. Although the 2-point statistics can be straightforwardly extended to the 3-point correlation function \cite{10.1046/j.1365-8711.2003.06321.x,Slepian_2017} and bispectrum \cite{10.1093/mnras/stv961,10.1093/mnras/stw2679,2020JCAP...03..040H,2021JCAP...04..029H,2024PhRvD.109h3534H}, as well as the 4-point correlation function \cite{2021arXiv210801670P,Philcox_2022} and trispectrum etc.\cite{Gualdi_2021,Gualdi_2022} for the recovery of the lost non-Gaussian information. These higher-order statistics are computationally expensive, with signal-to-noise ratios that degrade rapidly at higher orders, and have a large dimensionality, so the model of their uncertainties is difficult and also computationally intensive. The challenges imposed by the n-point statistics motivate the development of various novel summary statistics and the reinforcement of previous alternative clustering methods, including but not limited to: proxies of higher order n-point statistics \cite{2022arXiv221012743H,PhysRevD.91.043530,Dizgah_2020}, the marked 2-point correlation function \cite{White_2016,PhysRevD.97.023535,10.1093/mnras/sty1335} or power spectrum \cite{2021PhRvL.126a1301M,2022arXiv220601709M}, the density-split clustering \cite{10.1093/mnras/stab1654,Paillas_2023,2024MNRAS.531..898P,2024MNRAS.531.3336C}, the nearest neighbor distributions \cite{10.1093/mnras/staa3604,10.1093/mnras/stab961}, the void statistics \cite{Kreisch_2019,Massara_2015,10.1093/mnras/stv777,Hamaus_2015,Kreisch_2022,2019BAAS...51c..40P}, the counts-in-cells statistics \cite{2020MNRAS.495.4006U,10.1093/mnras/sty2802}, the minimum spanning tree \cite{2020MNRAS.491.1709N,Naidoo_2022}, and the wavelet scattering transform \cite{2021arXiv210807821V,Valogiannis_2022,Valogiannis_2024,2024arXiv240718647V}.

In our previous works, we have investigated the MFs as one of the interesting alternative statistics for the LSS. We found that the MFs of LSS (on nonlinear scales) are promising in probing departures from general relativity \cite{2017PhRvL.118r1301F} and detecting the signatures left by massive neutrinos \cite{2022JCAP...07..045L} (also see Liu et al. \cite{PhysRevD.101.063515}). Using the Fisher matrix formalism based on cosmological N-body simulations, we explored in a quantitive way the information content of the MFs (on nonlinear scales) and found the MFs can provide complementary information to the power spectrum and improve the constraints on cosmological parameters \cite{2021arXiv210803851J}, the sum of neutrino mass \cite{2022JCAP...07..045L,2023JCAP...09..037L}, and the modified gravity parameters \cite{2023arXiv230504520J}. However, on nonlinear (even quasi-linear) scales, there still lacks a reliable model to accurately predict the MFs for observed galaxies: the weakly non-Gaussian formulae derived by \cite{2003ApJ...584....1M,PhysRevD.104.103522} disagree with the real space particle distribution from the N-body simulation unless large smoothing scales are used \cite{2004astro.ph..8428N,PhysRevD.105.023527}; the effect of redshift-space distortions (RSD) \cite{1987MNRAS.227....1K,1972MNRAS.156P...1J,1998ASSL..231..185H} on the MFs is only accurately modeled for Gaussian \cite{1996ApJ...457...13M} and weakly non-Gaussian fields \cite{2013MNRAS.435..531C}; in addition, the weakly non-Gaussian MFs are bias-independent only when they are written as a function of the filling factor threshold\footnote{See Sec.4 of \cite{2023JCAP...09..037L} for a detailed discussion about the filling factor threshold and its difference from the threshold used in this work.} and the galaxy bias is local and monotonic \cite{2003ApJ...584....1M,2013MNRAS.435..531C}.

In this work, for the first time, we derive cosmological constraints from both the Gaussian and non-Gaussian information embedded in the MFs of observed galaxies, specifically the BOSS CMASS galaxies, based on a simulation-based forward model, where we model the non-linear structure growth and cosmological dependence of LSS with high-resolution cosmological N-body simulations, connect galaxies with dark matter halos by employing the halo occupation distribution (HOD) framework \cite{Zheng_2005,Zheng_2007}, and forward model the effects of redshift space distortions, Alcock–Paczynski (AP) distortions \cite{1979Natur.281..358A}, and observational systematics like survey geometry and selection functions. We adopt machine learning techniques to learn the MFs measured from mock galaxy catalogs where all ingredients are included. Finally, the forward model of the MFs is used to infer cosmological parameters and HOD parameters from the CMASS data.  We also compare with the 2-point correlation function (2PCF) forward-modeled with the same methodology, which serves as both a benchmark for the constraining power of the MFs and a cross-check of our forward model pipeline.

This paper is organized as follows. Section \ref{sec:cmass_sample} describes the CMASS sample of BOSS DR12 used in this work. We elucidate how to construct the simulation-based forward model in Section~\ref{sec:forward_model}. After validating our model with internal and external datasets in Section~\ref{sec:validate_emulator}, we present our main results in Section~\ref{sec:results} and discuss comparisons with previous works in Section \ref{sec:discussions}. Finally, we conclude in Section \ref{sec:conclusion}. We discuss some subtleties of our pipeline and present the constraints on HOD parameters in the appendices. 

\section{BOSS CMASS data}
\label{sec:cmass_sample}
\subsection{BOSS CMASS galaxy catalog}
We use the CMASS galaxy sample from Data Release 12 (DR12) of the Baryon Oscillation Spectroscopic Survey (BOSS) \cite{2013AJ....145...10D}, which is part of the SDSS-III program \cite{Eisenstein_2011}. The CMASS sample consists mainly of Luminous Red Galaxies (LRG) \cite{10.1093/mnras/stv2382}, and has very high completeness down to stellar mass of $M_* \approx 10^{11.3} \mathrm{M}_{\odot}$ for $z>0.45$ \cite{10.1093/mnras/stt1424}. Following \cite{Valogiannis_2022,Valogiannis_2024},  we measure the summary statistics for the galaxies in both the Northern (NGC) and Southern Galactic Cap (SGC) and denote them as $X_{\rm NGC}$ and $X_{\rm SGC}$, respectively. Our analysis will focus on the average of $X_{\rm NGC}$ and $X_{\rm SGC}$ weighted by the angular area of the northern and southern cap, $A_{\rm NGC}$ and $A_{\rm SGC}$,  
\begin{equation}
\label{eq:area_mean}
X_{\mathrm{N}+\mathrm{S}}=\frac{\left(A_{\mathrm{NGC}} X_{\mathrm{NGC}}+A_{\mathrm{SGC}} X_{\mathrm{SGC}}\right)}{\left(A_{\mathrm{NGC}}+A_{\mathrm{SGC}}\right)}.
\end{equation}
To reduce the influence of angular incompleteness on the MFs, we will focus on the angular regions where the completeness is higher than 0.9, as was done in \cite{2022ApJ...928..108A}. We refer readers to Figure~8 of \cite{10.1093/mnras/stv2382} for the visualization of the completeness map. Therefore, the used areas for the CMASS North and South samples are about $6912 ~\rm{deg}^2$ and $2558 ~\rm{deg}^2$, respectively.

\begin{figure}[tbp]
	\centering
	\includegraphics[width=1.0\textwidth]{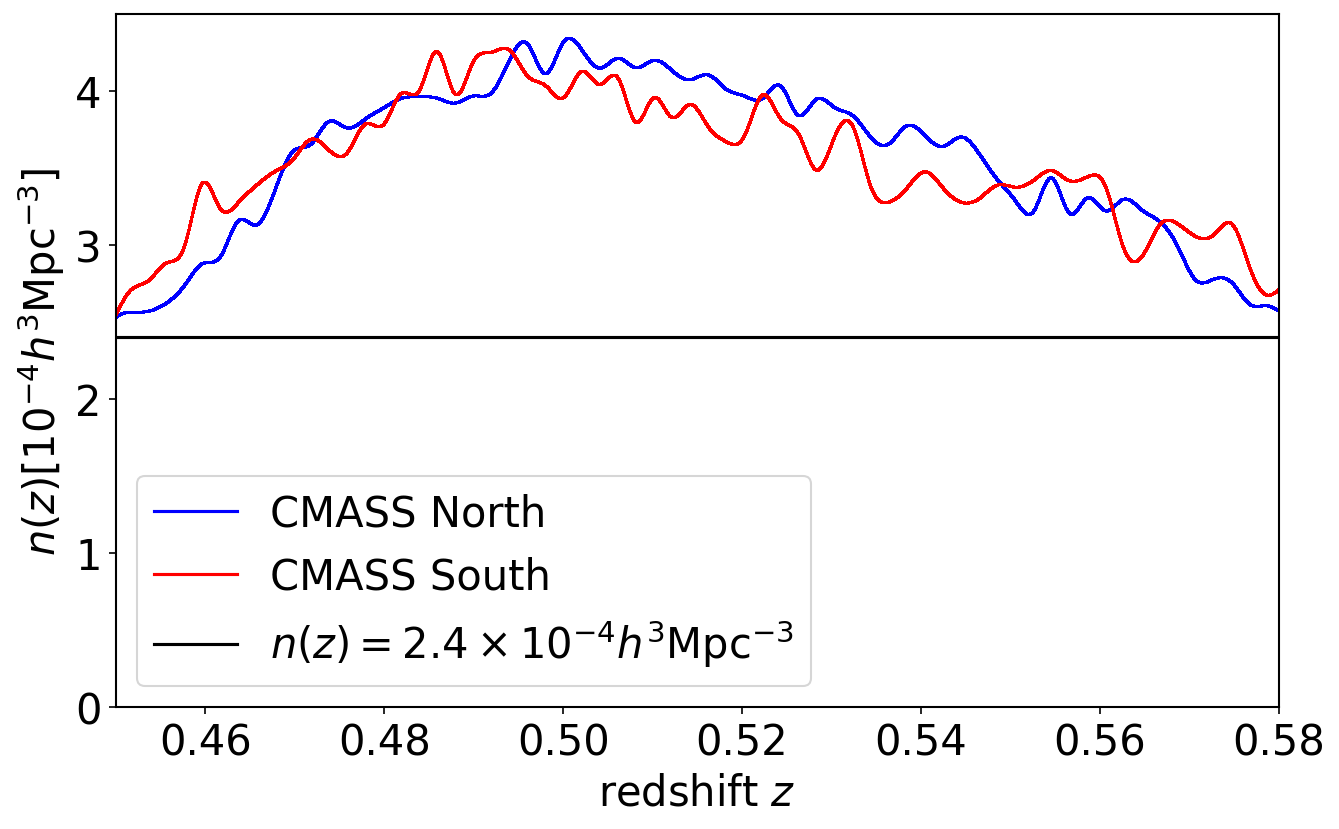}
	\caption{\label{fig:CMASS_North_South_Nz} The galaxy number density of the CMASS North (blue) and South (red) samples as a function of redshift. Both samples are downsampled to have a constant number density $n(z)=2.4\times 10^{-4}~h^3\rm{Mpc}^{-3}$, which is chosen to be the minimum number density of the Patchy mocks in this redshift range and is
slightly smaller than the optimal choice $n(z)=2.5\times 10^{-4}~h^3\rm{Mpc}^{-3}$ for the CMASS sample.  All mock galaxy catalogs used in this work are downsampled to match this constant number density.} 
\end{figure}

We will explore the redshift range $0.4<z<0.7$ and attempt to find a subsample of CMASS galaxies with a high number density and a large survey volume at the same time.  Each galaxy is first weighted with the total systematic weight suggested in \cite{10.1093/mnras/stv2382}:
\begin{equation}
\label{eq:sys_tot}
w_{\rm sys,tot}=w_{\rm sys}(w_{\rm fc}+w_{\rm zf}-1),
\end{equation}
where the weights $w_{\rm sys}$, $w_{\rm fc}$, and $w_{\rm zf}$ are used to correct imaging systematics, fiber collisions, and redshift failures. The galaxies are then binned into redshift shells of thickness $\Delta z = 0.002$. The volume of each shell is calculated assuming a flat $\Lambda$CDM cosmology with a matter density parameter $\Omega_m=0.3152$ and a dimensionless Hubble parameter $h=0.6736$, this is our fiducial cosmology and corresponds to cosmology c000 which will be described in details in Sec~\ref{sec:simulations}. We decide to work with the galaxies in the redshift range $0.45<z<0.58$, whose number density is higher than $\bar{n}\sim 2.5\times 10^{-4} h^{3}\rm{Mpc}^{-3}$, as shown in figure~\ref{fig:CMASS_North_South_Nz}. The galaxies in each redshift bin are then randomly downsampled to have a constant galaxy number density $\bar{n}=2.4\times 10^{-4} h^{3}\rm{Mpc}^{-3}$. The redshift bin thickness varies from $4.4~h^{-1}\rm{Mpc}$ to $4.7~h^{-1}\rm{Mpc}$, it is much smaller than the scale used to smooth the density field. Hence, we anticipate this choice of redshift bin thickness won't impact our results significantly. With the above steps done for both NGC and SGC, we obtain a sample with a total volume of $\sim  (1.2~h^{-1}\rm{Gpc})^3$, and an effective volume of $\sim (0.9~h^{-1}\rm{Gpc})^3$, which is calculated with 
\begin{equation}
V_{\text {eff }}=\int\left[\frac{\bar{n} P_0}{1+\bar{n} P_0}\right]^2  \mathrm{~d} V(z) = \left[\frac{\bar{n} P_0}{1+\bar{n} P_0}\right]^2 V,
\end{equation}
where $P_0=10000h^{-3}\rm{Mpc}^3$, and the second equality holds because the number density of galaxies is kept constant in the redshift range.

We focus on the CMASS catalog rather than both the LOWZ and CMASS catalogs because their effective redshifts differ. The effective redshift for the CMASS catalog is approximately $0.5$, while for the LOWZ catalog, it is around $0.3$. In our forward model, we aim to match the effective redshift of the CMASS data by using simulation outputs at redshift $z=0.5$, which require over 16 TB of disk space. Given the large size of the simulation dataset, downloading and analyzing the data is time-consuming. To include the analysis of the LOWZ catalog, we would need to download simulation outputs at a redshift close to $z=0.3$ and repeat the forward modeling pipeline. While we recognize the relevance of the LOWZ catalog, this work represents our first application of the MFs to observed galaxies using the simulation-based method. We plan to explore the inclusion of the LOWZ catalog in future research.

\subsection{MD-Patchy mocks}
\label{sec:patchy_mocks}
We use the MultiDark-Patchy mocks (Patchy mocks, hereafter) \cite{2016MNRAS.456.4156K,10.1093/mnras/stw1014}, which consists of 2048 mock galaxy catalogs, to estimate the covariance matrix of the data vector.  These mocks are generated following multiple procedures to reproduce the number density, selection function, survey geometry, two-point statistics, and three-point statistics of BOSS DR11$\&$DR12 galaxy samples. The first step is the construction of mock galaxy catalogs in periodic boxes using the PATCHY code \cite{10.1093/mnrasl/slt172,2015MNRAS.450.1836K}  based on Augmented Lagrangian Perturbation Theory and a bias scheme,  where the bias is tuned to fit the proper clustering of the BigMultiDark Planck simulation \cite{2016MNRAS.457.4340K} that matches the BOSS LRG clustering. The BigMultiDark Planck simulation is run using the GADGET-2 code \cite{Springel_2005} with a cosmology specified by $\left\{\Omega_{\rm b}, \Omega_m, \sigma_8, n_s, h\right\}=\{0.0482, 0.307, 0.829, 0.961, 0.6778\}$, it follows the evolution of $3840^3$ dark matter particles in a volume of $(2.5~h^{-1}\rm{Gpc})^3$. A mock galaxy catalog is then constructed from the halo catalog of this simulation using the Halo Abundance Matching technique \cite{2004ApJ...609...35K} to reproduce the clustering of the BOSS DR11$\&$DR12 galaxies.  Finally,  the sugar code \cite{10.1093/mnras/stw1014} is applied to build the light cones with the model of observational effects such as incompleteness, geometry, veto masks, and fiber collisions. We will use the weights provided by the Patchy mocks 
\begin{equation}
\label{eq:weight_patchy}
w_{\rm tot}=w_{\rm fc}w_{\rm veto},
\end{equation}
to correct the effect of fiber collisions and remove galaxies excluded by veto masks.

\section{The simulation-based forward model }
\label{sec:forward_model}
\subsection{The \textsc{Abacus}\textsc{Summit} simulations}
\label{sec:simulations}

In this work, we use the \textsc{Abacus}\textsc{Summit} simulations \cite{2021MNRAS.508.4017M} to model the cosmological dependence of the MFs. \textsc{Abacus}\textsc{Summit} is a suite of cosmological N-body simulations designed to meet and exceed the requirements of the Dark Energy Spectroscopic Instrument (DESI) survey \cite{2013arXiv1308.0847L}.  The simulation suite is run using the Abacus N-body code \cite{10.1111/j.1365-2966.2005.09655.x,10.1093/mnras/stab2482}, the base simulations follow the evolution of $6912^3$ CDM particles in a cosmological volume of $\left(2~h^{-1} \mathrm{Gpc}\right)^{3}$, while the small-box simulations (denoted as \textsc{Abacus}\textsc{Small} hereafter) evolve $1728^3$ CDM particles in a $\left(0.5~h^{-1} \mathrm{Gpc}\right)^{3}$ box.

We will use a subset of the \textsc{Abacus}\textsc{Summit} simulation suite with 85 different cosmologies, the eight cosmological parameters $\left\{ \omega_{\rm b}, \omega_{\mathrm{cdm}}, \sigma_8, n_s, \alpha_s, N_{\mathrm{eff}}, w_0, w_a\right\}$ vary from one cosmology to another, where $\omega_{\mathrm{b}}=\Omega_{\rm b} h^2$ and $\omega_{\mathrm{cdm}}=\Omega_c h^2$ are the physical baryon and cold dark matter densities, $\sigma_8$ is the amplitude of linear density fluctuations at $8~h^{-1}\rm Mpc$, $n_s$ is the spectral index, $\alpha_s$ is the running of the spectral index, $N_{\mathrm{eff}}$ is the effective number of ultra-relativistic species, $w_0$ and $w_a$ are the two parameters controlling the time-varying dark energy equation of state by $w(a)=w_0+w_a(1-a)$ and $a$ is the scale factor. The value of the Hubble constant $H_0$ is chosen to match the sound horizon at last scattering $\theta_{*}$ derived from Planck 2018 measurement \cite{2020A&A...641A...6P}, and a flat spatial curvature is assumed for all cosmologies.  Specifications of the \textsc{Abacus}\textsc{Summit} simulations used in this work are listed below \footnote{We use the same naming scheme cXXX, where XXX goes from 000 to 181, as that used in \cite{2021MNRAS.508.4017M} for different cosmologies.}:

c000:  The fiducial cosmology, with parameter values matched with the mean estimates of the TT, TE, EE+lowE+lensing likelihood chains for the Planck 2018 results \cite{2020A&A...641A...6P}, we also use it as our fiducial cosmology in this work. There are 25 realizations for the base simulations and 1883 realizations for the small-box simulations. They will be used to estimate the covariance matrix of statistics.

c001-004: Four secondary cosmologies, including a low $\omega_{\rm cdm}$ model corresponding to WMAP9 + ACT + SPT \cite{PhysRevD.95.063525}, a thawing dark energy ($w$CDM) model, a high $N_{\rm eff}$ model, and a low $\sigma_8$ model.

c013: A cosmology that corresponds to $Euclid$  Flagship2 \cite{2024arXiv240513495E} $ \Lambda$CDM run.

c100-126: A series of cosmologies that vary the eight cosmological parameters one by one with both negative and positive steps centered at the fiducial cosmology.

c130-181: 52 cosmologies to form an emulator grid and cover the region $3-8$ standard deviations beyond current constraints from the combination of CMB and LSS data in the 8D cosmological parameter space.

Together with the first realization of c000, the other 84 cosmologies used in this work share the same phase seed. The parameter ranges of the 85 cosmologies can be found in Table \ref{tab:models}. More details of each cosmology are provided on the \textsc{Abacus}\textsc{Summit} website \footnote{\href{https://abacussummit.readthedocs.io/en/latest/abacussummit.html}{https://abacussummit.readthedocs.io/en/latest/abacussummit.html}}. The distribution of the 85 cosmologies is visualized in Fig.~1 of \cite{2022MNRAS.515..871Y}, where the five cosmologies c000-004 are highlighted as the test set, we will also follow this choice.

\subsection{The halo-galaxy connection model}
\label{sec:halo_galaxy_connection}
The halos are found on-the-fly using a new specialized spherical-overdensity-based halo finder dubbed COMPASO \cite{10.1093/mnras/stab2980}. We populate halos with mock galaxies using the Halo Occupation Distribution framework based on the halo catalogs at $z=0.5$. To efficiently generate galaxy mocks from the \textsc{Abacus}\textsc{Summit} simulations (the Abacus mocks, hereafter), we use the highly optimized AbacusHOD implementation\footnote{The code is publicly available as a part of the ABACUSUTILS package at https:// github.com/abacusorg/abacusutils. Example usage can be found at https://abacusutils.readthedocs. io/en/latest/hod.html.} \cite{10.1093/mnras/stab3355}, which adopts the specific parametrization well-suited model for LRG \cite{Kwan_2015}:
\begin{equation}
\label{eq:n_cen}
\bar{n}_{\mathrm{cent}}(M)=\frac{1}{2} \operatorname{erfc}\left[\frac{\log _{10}\left(M_{\mathrm{cut}} / M\right)}{\sqrt{2} \sigma}\right],
\end{equation}
\begin{equation}
\label{eq:n_sat}
\bar{n}_{\mathrm{sat}}(M)=\left[\frac{M-\kappa M_{\mathrm{cut}}}{M_1}\right]^\alpha \bar{n}_{\mathrm{cent}}(M),
\end{equation}
where $M_{cut}$ determines the minimum halo mass to host a central galaxy, $M_1$ sets the typical halo mass that hosts one satellite
galaxy, $\sigma$ characterizes the slope of the transition from 0 to 1 in the number of central galaxies, $\alpha$ is the power-law index for the number of satellite galaxies, $\kappa M_{cut}$ specifies the minimum mass required to host a satellite. In Eq~\ref{eq:n_sat}, the $\bar{n}_{\mathrm{cent}}$ term is used to remove satellites from haloes without central galaxies. 

We have also implemented the velocity bias for both centrals $\alpha_{\rm vel,c}$ and satellites $\alpha_{\rm vel,s}$, where $\alpha_{\rm vel,c}$ describes the difference in the peculiar velocity between the central galaxy and the halo center, while $\alpha_{\rm vel,s}$ characterizes the deviation of satellite galaxies' peculiar velocities from those of local dark matter particles. When $\alpha_{\rm vel,c} = 0$ and $\alpha_{\rm vel,s} = 1$, there is no velocity bias for both central and satellite galaxies. The velocity bias is a necessary extension of the HOD framework since evidence supporting the existence of velocity bias has been found in the analysis of both hydrodynamical simulations \cite{Ye_2017,10.1093/mnras/stac830} and observational galaxy catalogs \cite{10.1093/mnras/stu2120,10.1093/mnras/stab3355}. The velocity bias for both central and satellite galaxies must be included to explain both the Finger-of-God (FoG) effect at non-linear scales and the Kaiser effect in the linear regime, seen in the two-point correlation function of the CMASS galaxies \cite{10.1093/mnras/stu2120}. Whether it can also faithfully explain the RSD effect, especially the FoG effect on small scales, detected by non-Gaussian statistics is an interesting question that warrants further investigation.

In summary, we model the halo-galaxy connection with an extended HOD framework parameterized by a total of 7 parameters: $\{ M_{\rm cut}, M_1,\sigma,\alpha,\kappa,\alpha_{\rm vel,c},\alpha_{\rm vel,s} \}$, 
a short description of the meaning and the value range for each parameter are listed in table~\ref{tab:models}.  To obtain good representatives of the 7D HOD parameter space, we generate a Latin hypercube \cite{ef76b040-2f28-37ba-b0c4-02ed99573416} with 42500 samples and assign 500 HOD variations to each of the 85 cosmologies for the training and test dataset. Another $125\times85$ Latin hypercube samples are generated for the validation and test dataset, they are also evenly distributed to each of the cosmologies. Our choice of the number of HOD variations is in between the 100 variations used in \cite{2024MNRAS.531.3336C} and the 2700 variations adopted in \cite{Valogiannis_2024}, this can lead to a difference in the emulator error. We find a considerable improvement in the constraining power of the summary statistics can be made if the emulator error is significantly reduced in Appendix~\ref{sec:pipetest}.

\begin{center}
	\small
	\begin{table}[tbp]
	\scalebox{0.9}[1]{
    	\begin{tabular}{|c|c|c|}
            \hline
            Parameter & Meaning & Range \\
            \hline
              $\omega_{\rm b}$ & Physical baryon  density & {$[0.0207,0.0243]$} \\
             $\omega_{\mathrm{cdm}}$ & Physical cold dark matter density & {$[0.103,0.140]$} \\
             $\sigma_8$ & Amplitude of the linear power spectrum at $8 h^{-1}\mathrm{Mpc}$ & {$[0.678,0.938]$} \\
             $n_s$ & Spectral index of the primordial power spectrum & {$[0.901,1.025]$} \\
             $\alpha_s$ & Running of the spectral index & {$[-0.038,0.038]$} \\
             $N_{\text {eff}}$ & Effective number of relativistic species & {$[2.1902, 3.9022]$} \\
             $w_0$ & Dark energy equation-of-state at $z=0$ & {$[-1.27,-0.70]$} \\
             $w_a$ & Time evolution of dark energy equation-of-state & {$[-0.628,0.621]$} \\
            \hline
             $\log _{10} M_{\text {cut }}$ & Typical mass to host a central & {$[12.4,13.3]$} \\
             $\log _{10} M_1$ & Typical mass to host one satellite & {$[13.0,15.0]$} \\
             $\log _{10} \sigma$ & Turn on slope for central occupation & {$[-3.0,0.0]$} \\
             $\alpha$ & Power-law index for the mass dependence of the number of satellites & {$[0.5,1.5]$} \\
             $\kappa$ & Parameter defining the minimum mass to host a satellite & {$[0.0,8.0]$} \\
             $\alpha_{\text {vel, c }}$ & Central velocity bias & {$[0.0,0.8]$} \\
             $\alpha_{\text {vel, s }}$ & Satellite velocity bias & {$[0.0,1.5]$} \\
            \hline
        \end{tabular}
        }
    \caption{\label{tab:models} The eight cosmological parameters and seven HOD parameters used in our emulators. We list the parameter symbols, their physical meanings, and their ranges in the training set. }
    \end{table}
\end{center}

\subsection{Survey systematics}
\label{sec:survey_systematics}
It is vital to understand all kinds of systematics existing in the redshift surveys, a part of them can be corrected or reduced before comparing with mock galaxy catalogs. We rely on the weights suggested in \cite{10.1093/mnras/stv2382} to account for imaging systematics, redshift failures, and fiber collisions. The angular selection effect is alleviated by cutting out angular regions with completeness smaller than 0.9, while the radial selection effect is controlled by downsampling the galaxies to have a constant number density in the chosen redshift range. In addition, we use a relatively large smoothing scale to further reduce the effects of these systematics. 

It is equally important to forward model the other part of systematics as realistically as possible in the mock galaxy catalogs so that the difference in the summary statistics between the observed galaxies and the mock galaxies is mainly from the variations in cosmological and HOD models instead of the remaining systematics. 

\begin{figure}[tbp]
	\centering
	\includegraphics[width=1.0\textwidth]{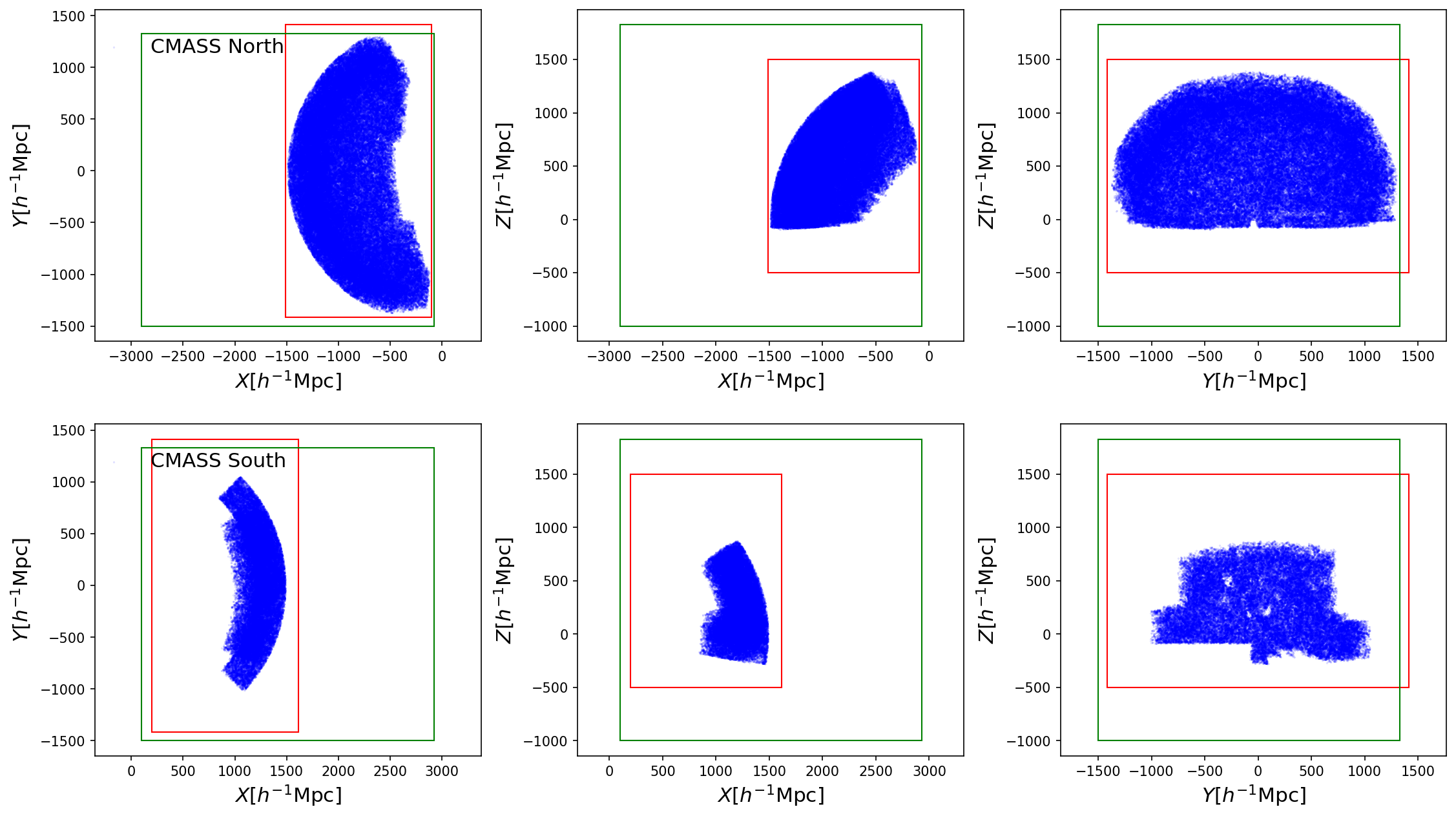}
	\caption{\label{fig:ns_cutsky_xyz} The 3D spatial distribution of the Abacus mock galaxies produced with the pipeline described in the text for the CMASS North (first row) and South (second row) samples. The position of mock galaxies is projected along the $\hat z$ (first column), $\hat y$ (first column), $\hat x$ (first column) axis. The red rectangles outline the remapped simulation box, while the green squares delineate the cubic grid where the number density field is constructed from the positions of galaxies.} 
\end{figure}

Start with catalogs of galaxies in a periodic simulation box with a volume of $(2~h^{-1}\rm{Gpc})^3$, we first remap the simulation box to a cuboid with dimensions $1.414\times 2.828\times 2.0~(h^{-1}\rm{Mpc})^3$ using the volume remapping method \cite{2010ApJS..190..311C} so that the survey geometry for the CMASS sample can be efficiently fit. The remappings are one-to-one, volume-preserving, and keep local structures intact, which means the method does not introduce extra systematics for statistics like the MFs. Then, we rotate and translate the cuboid to a proper place so that the CMASS NGC and SGC galaxy samples in the redshift range $0.45<z<0.58$ can both be fully embedded in the cuboid. To help understand how this is done, we plot in figure~\ref{fig:ns_cutsky_xyz} the cuboid together with the mock galaxy catalog that will be obtained with the following operations for both CMASS North and South samples. Using the true underlying cosmology for mock galaxies of the 85 cosmologies, we convert the comoving Cartesian coordinates ($x,\ y,\ z$) of mock galaxies into Sky coordinates (right ascension (RA), declination (DEC), and redshift $z$) and then apply the redshift distortions to each galaxy. After removing galaxies with redshift $z<0.45$ or $z>0.58$, we further discard galaxies outside the survey geometry and galaxies inside the DR12 veto masks, including the Centerpost mask, collision priority mask, bright stars mask, bright objects mask, bad field mask, non-photometric conditions mask, seeing cut mask, and extinction cut mask. We refer readers to Sec~5.1.1 of \cite{10.1093/mnras/stv2382} for a detailed description of these masks. Lastly, we follow what has been done for the BOSS CMASS galaxy catalog in Sec~\ref{sec:cmass_sample} to reduce the impact of angular and radial selection effects: for the former, galaxies are discarded in angular regions where the completeness is lower than 0.9; for the latter, galaxies are binned into redshift shells and downsampled have a constant number density $\bar{n}=2.4\times 10^{-4} h^{3}\rm{Mpc}^{-3}$ throughout the redshift range $0.45<z<0.58$, the number density is calculated using the fiducial cosmology. 

The possible inconsistency between our fiducial cosmology and the underlying cosmology can lead to the so-called Alcock–Paczynski (AP) distortion. To forward model the AP effect, we always adopt the fiducial cosmology for the Sky-to-Cartesian coordinate conversion of all mock galaxy catalogs. The forward-modeled galaxy catalog is shown in figure~\ref{fig:ns_cutsky_xyz} for both CMASS North and South samples. We note a large fraction of the remapped simulation box is not used by both the north and south samples. To increase the usage of the simulation volume, we rotate the simulation box before the volume remapping step and repeat the procedure described in the previous paragraph three times. The summary statistics measure for the north and south samples are first combined with Eq~\ref{eq:area_mean} and then averaged over the three rotations.

\subsection{The summary statistics}
\label{sec:tpcf_mf}
We describe how to measure the Minkowski functionals and two-point correlation function for the CMASS galaxy catalogs and mock galaxy catalogs constructed from cosmological simulations. 

\subsubsection{Minkowski functionals}
\label{sec:minkowski}
For a scalar random field $u$ defined on a Euclidean and three-dimensional manifold $\mathbb{M}$, it is a usual practice to study its geometry and topology by investigating these properties for the excursion set defined as 
\begin{equation}
E_\nu=\{\boldsymbol{x} \in \mathbb{M}: u(\boldsymbol{x}) \geqslant \nu\},
\end{equation}
where $\nu$ is the threshold parameter. When working with the periodically tiled Euclidean grid, like cosmological simulation boxes, the volume-normalized Minkowski functionals can be written as
\begin{equation}
\label{eq:mfs_def}
\begin{gathered}
W_0=\frac{1}{V} \int_{E_\nu} d V \\
W_1=\frac{1}{6 V} \int_{\partial E_\nu} d A \\
W_2=\frac{1}{6 \pi V} \int_{\partial E_\nu} k_1+k_2 d A \\
W_3=\frac{1}{4 \pi V} \int_{\partial E_\nu} k_1 k_2 d A
\end{gathered}
\end{equation}
where $k_1$ and $k_2$ are the principal curvatures for the surface of the excursion set $\partial E_\nu$, $dV$ and $dA$ are the infinitesimal volume and area element. 

To help understand the information content embedded in the MFs, we cite here the analytic formula of the Minkowski functionals (up to first order) for the weakly non-Gaussian field in real space \cite{2003ApJ...584....1M}
\begin{equation}
\label{eq:first_nonGaussian}
\begin{aligned}
W_k(\nu)= & \frac{1}{(2 \pi)^{(k+1) / 2}} \frac{\omega_3}{\omega_{3-k} \omega_k}\left(\frac{\sigma_1}{\sqrt{3} \sigma_0}\right)^k \\
& \times e^{-\nu^2 / 2}\left\{H_{k-1}(\nu)+\left[\frac{1}{6} S^{(0)} H_{k+2}(\nu)+\frac{k}{3} S^{(1)} H_k(\nu)+\frac{k(k-1)}{6} S^{(2)} H_{k-2}(\nu)\right] \sigma_0\right\}.
\end{aligned}
\end{equation}
$\omega_0=1$, $\omega_1=2$, $\omega_2=\pi$, $\omega_3=4\pi/3$, and the parameters $\sigma_j$ are given by 
\begin{equation}
\sigma_j^2=\int_0^\infty \frac{k^2 dk}{2\pi^2}k^{2j}P_L(k)e^{-(kR)^2},
\end{equation}
where $P_L(k)$ is the linear power spectrum and the Gaussian window function is assumed. $H_{k}(\nu)$ is the $k$th order Hermite polynomials. $S^{(0)}$, $S^{(1)}$, and $S^{(2)}$ are three skewness parameters\footnote{See \cite{2003ApJ...584....1M} for the definition of the skewness parameters}.

Previous application of the MFs to the CMASS data \cite{Appleby_2020,2022ApJ...928..108A} focused on the amplitude of the MFs, which is determined by the ratio $\sigma_1/\sigma_0$ and only Gaussian information is extracted. In real space, the amplitude of the MFs provides a measure of the shape of the linear matter power spectrum and hence is mainly sensitive to $\omega_{\rm cdm}$ and $n_s$. In redshift space, the amplitude is changed by redshift space distortions and influenced by the linear growth rate and galaxy bias. We adopt a density contrast threshold parameter $\delta$ in this analysis, which is related to the one used in Eq~\ref{eq:first_nonGaussian} with $\delta=\nu \sigma$ ($\sigma$ is the standard deviation of the smoothed density contrast field). Therefore, any factor, physical or unphysical, that changes the variance of the density field will rescale the threshold $\delta$, and the curves of the MFs as a function of $\delta$ will be expanded or compressed. This choice of threshold parameter allows us to extract more information at the cost of being prone to systematics. Most importantly, our full shape analysis of the MFs allows us to probe beyond Gaussian information in the MFs. As seen in Eq~\ref{eq:first_nonGaussian}, the departure from the Gaussian prediction of the MFs is sensitive to the skewness parameters. For a non-Gaussian field like what is studied in this work, the MFs also depend on the kurtosis, and even higher-order parameters \cite{2004astro.ph..8428N,PhysRevD.105.023527}. The constraining power of the MFs on cosmological and HOD parameters comes from the combination of both Gaussian and non-Gaussian information embedded in the MFs.

For galaxy samples with non-trial boundaries seen in figure~\ref{fig:ns_cutsky_xyz}, we adopt a similar algorithm to that used in \cite{2022ApJ...928..108A} to obtain an unbiased estimate of the MFs. The galaxies are first embedded into a uniform cubic grid of size $L_{\rm box}=2828~h^{-1}\rm Mpc$ and spacing $\Delta_g=2828/512\simeq 5.5~h^{-1}\rm Mpc$ for both the North and South sample. Then, we interpolate the positions of galaxies onto the cubic grid using the Cloud-in-Cell (`CIC') \footnote{We use the routine provided by Pylians (Python libraries for the analysis of numerical simulations \cite{2018ascl.soft11008V}): https://pylians3.readthedocs.io/en/master/, other mass assignment schemes are also available, such as `NGP' (nearest grid point), `TSC' (triangular-shape cloud), and 'PCS' (piecewise cubic spline)} mass assignment scheme with the weight associated to each galaxy (uniform weight is used for galaxy catalogs apart from the CMASS sample and the Patchy mocks) and get a number density field $n(\boldsymbol{x})$. The survey geometry and veto masks are projected into a 3D constant field $M(\boldsymbol{x})$ with the same size and resolution as that used for the construction of the number density field, the value of $M(\boldsymbol{x})$ is set to 1 for regions inside the survey geometry but outside the veto masks, and set to 0 otherwise. For the regions with $M(\boldsymbol{x})=1$, we project the value of the angular completeness map onto them so that the 3D constant field now becomes a completeness field. 

The number density field and completeness field are both smoothed with a Gaussian filter $W\left(k R_{\mathrm{G}}\right) \propto \exp \left[-k^2 R_{\mathrm{G}}^2 / 2\right]$, we choose to work with $R_G=15~h^{-1}\rm{Mpc}$ and leave the investigation of multiple smoothing scales for future work. This smoothing scale is close to the mean galaxy separation $\bar{d} \sim 16~h^{-1}\rm{Mpc}$, hence this choice is aligned with previous analyses of observational data using the MFs or the genus alone\footnote{the genus $g$ is related with the Euler characteristic $V_3$ by $g=1-V_3$ \cite{1997ApJ...482L...1S}}, the choice of $R_G=\bar{d}/\sqrt{2}$ \cite{1994ApJ...420..525V,1989ApJ...340..625G} or $R_G=\bar{d}$ \cite{10.1093/pasj/55.5.911,2005ApJ...633....1P} is frequently used. We can in principle explore down to the resolution scale of the simulations with our simulation-based method. However, it is difficult to faithfully forward model all observational systematics on very small scales. As we mentioned in Sec~\ref{sec:survey_systematics}, we rely on the weights suggested in \cite{10.1093/mnras/stv2382} to account for imaging systematics, redshift failures, and fiber collisions. Our main concern is that the weight used to correct for the fiber collision effect in the CMASS galaxies may not be effective on small scales \cite{2017MNRAS.467.1940H}. We believe a comprehensive study of the impact of fiber collisions on higher-order statistics like the MFs, and how to properly correct or forward model this effect, is essential before going to smaller scales. However, this is beyond the scope of the current paper and is an area we are actively investigating. On the other hand, we find using a smaller scale doesn't necessarily guarantee tighter parameter constraints for the CMASS galaxy sample due to the low number density ($n(z)=2.4\times 10^{-4}~h^3\rm{Mpc}^{-3}$) adopted in this analysis. The larger shot noise on smaller scales can make the curves of the MFs more noisy, which leads to a lower accuracy of the emulator. The DESI LRG sample \cite{2024arXiv241112020D} has roughly twice this number density, allowing for the extraction of more non-Gaussian information on smaller scales.

To reduce the impact of complex boundaries and focus on regions with high completeness, we further discard regions where $M(\boldsymbol{x})<0.9$ after smoothing, which reduces the sample volume from $\sim (1.2~h^{-1}\rm{Gpc})^3$ to $\sim (1.0~h^{-1}\rm{Gpc})^3$. We then calculate the average number density $\bar{n}$ of galaxies and construct a density contrast field $\delta(\boldsymbol{x}) = n(\boldsymbol{x})/\bar{n}-1$, within the remaining volume.

We use Crofton's formula derived in \cite{1997ApJ...482L...1S} to numerically calculate the MFs for the excursion set constructed from the density contrast field. The thresholds are chosen based on the average MFs of the 2048 Patchy mocks with the same binning scheme as that used in our previous work \cite{2023JCAP...09..037L}: $N_b$ evenly spaced threshold bins are taken for $W_0$ in the range between $\delta_{0.01}$ to $\delta_{0.99}$, where the thresholds $\delta_{0.01}$ and $\delta_{0.99}$ correspond to the area (volume) fraction of 0.01 and 0.99, respectively. For other MFs, we first find the lower end of thresholds approximately corresponding to $1\%$ of the maximum of the statistics, $\delta_{low}$, and then find the higher end of thresholds where the curve is close to $1\%$ of the maximum of the statistics as well, $\delta_{high}$. Finally, $N_b$ evenly spaced threshold bins are taken in the range between $\delta_{low}$ to $\delta_{high}$. This threshold binning scheme tries to cover the variation range for each order of the statistics as extensively as possible while avoiding including the bins susceptible to noises and systematics. In this analysis, we take $N_b=60,\ 80,\ 100,\ 120$ for $W_0,\ W_1,\ W_2,\ W_3$ and concatenate the four MFs into a vector of length 360.

\begin{figure}[tbp]
	\centering
	\includegraphics[width=1.0\textwidth]{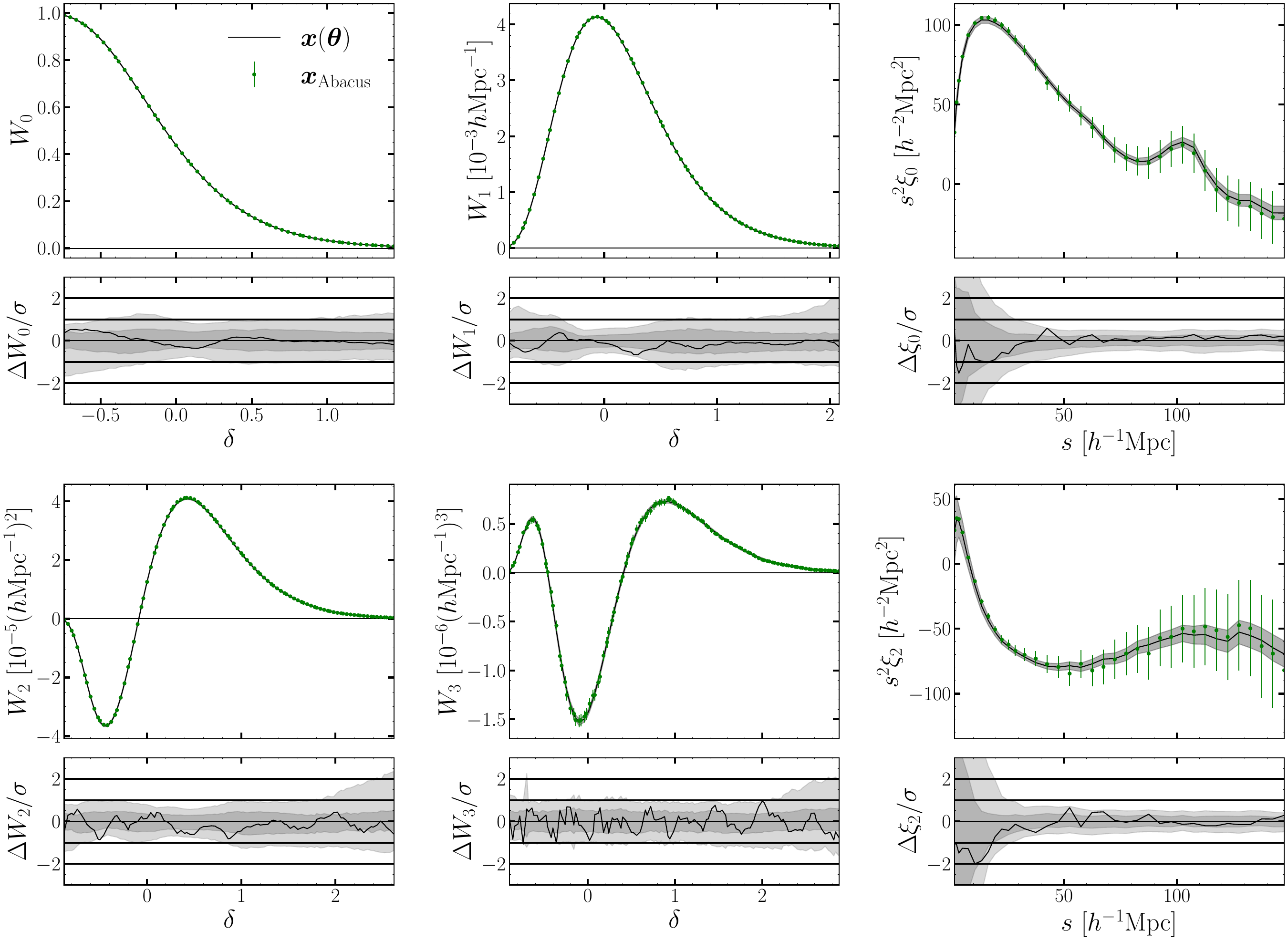}
	\caption{\label{fig:Emu_vs_cal} The comparison of the four MFs (first two columns) and the 2PCF (third column) between measurements from an Abacus mock and emulator predictions given the true parameter value of this mock. The first two MFs and the monopole moment of the 2PCF are displayed in the first row, while the last two MFs and the quadrupole moment are shown in the second row. The green points represent the measurements for the MFs or the 2PCF, the corresponding error bars are estimated with the 2048 Patchy mocks. The solid black lines show emulator predictions for the MFs or the 2PCF, and the semitransparent shadowed regions visualize the emulator error.  The difference between the emulator model and the data is also plotted in units of the data errors, where the $1\sigma$ and $2\sigma$ emulator errors are visualized with shaded regions while the $1\sigma$ and $2\sigma$ data errors are highlighted with horizontal black lines, respectively.} 
\end{figure}

\begin{figure}[tbp]
	\centering
	\includegraphics[width=1.0\textwidth]{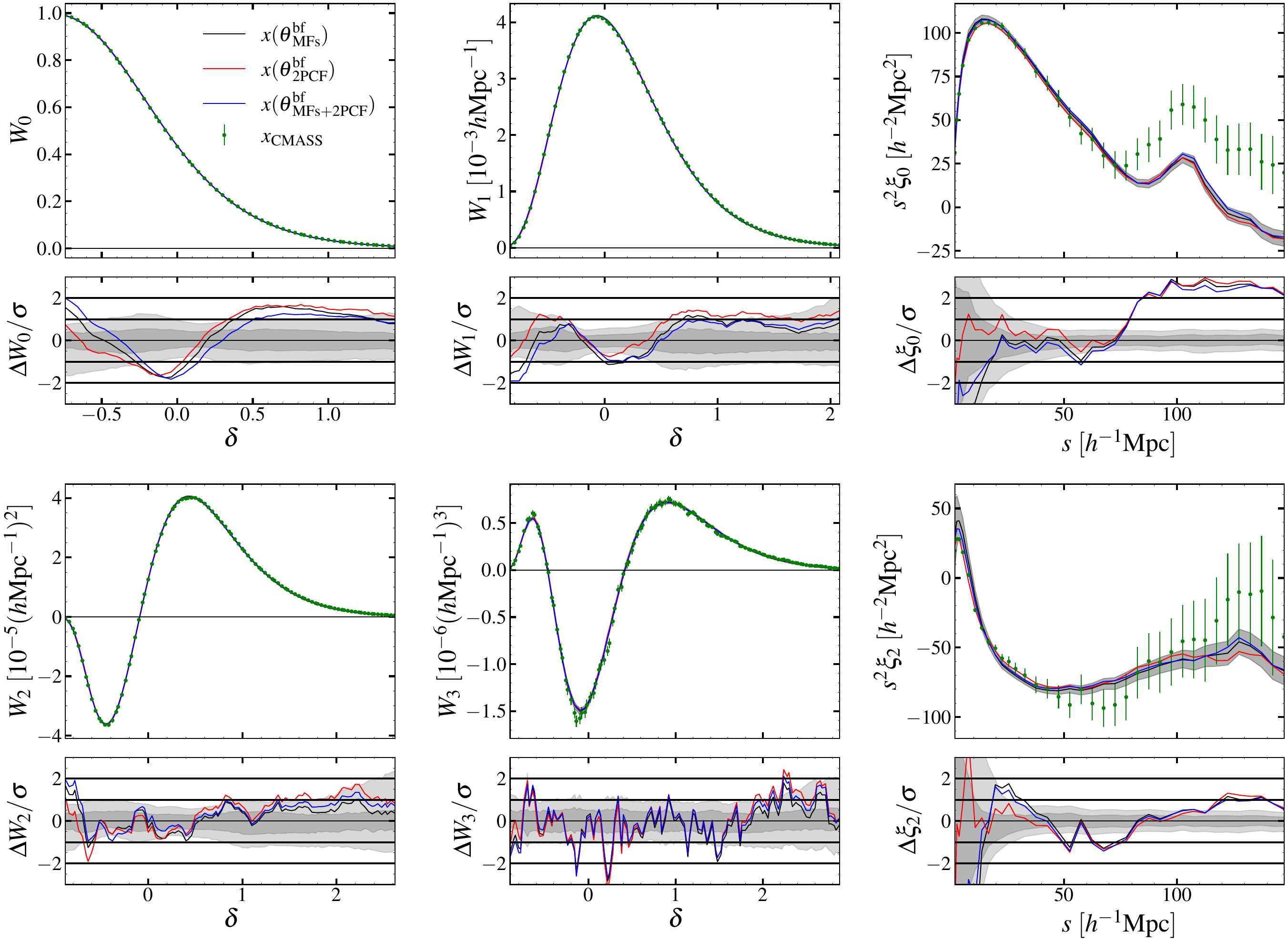}
	\caption{\label{fig:Emu_vs_CMASS} Similar to figure~\ref{fig:Emu_vs_cal}, but now the green points are the measured MFs or 2PCF from the CMASS galaxy sample. Three different kinds of emulator predictions are plotted here: $\boldsymbol{x}(\boldsymbol{\theta}^{\rm bf}_{\rm MFs})$, $\boldsymbol{x}(\boldsymbol{\theta}^{\rm bf}_{\rm 2PCF})$, and $\boldsymbol{x}(\boldsymbol{\theta}^{\rm bf}_{\rm MFs+2PCF})$  are the predictions of the statistics when inputting the best-fit value from the MFs, the 2PCF, and the joint combination of the MFs and 2PCF, which are plotted with black, red, and blue lines, respectively.} 
\end{figure}

We show the measurements of MFs for an Abacus mock and the CMASS sample in figure~\ref{fig:Emu_vs_cal} and figure~\ref{fig:Emu_vs_CMASS}, respectively. The MFs themselves and the error bars are shown with the predictions from the emulator and its inherent error. We will give a detailed interpretation of the two figures detailedly in Sec~\ref{sec:emulator_accuracy} and Sec~\ref{sec:cmass_fits}.

\subsubsection{2-point correlation function}
\label{sec:2ptpcf}
Following \cite{2024MNRAS.531.3336C,2024MNRAS.531..898P,Valogiannis_2024}, we use the two-point correlation function (2PCF) as a benchmark to assess the constraining power of the MFs. To capture the anisotropies introduced by redshift-space distortions and Alcock-Paczynski distortions we measure the correlation function as a function of the redshift space separation $s$ and the cosine of the angle between the redshift space separation vector and the line of sight $\mu$. The Landy $\&$ Szalay estimator \cite{1993ApJ...412...64L} is used to compute the 2PCF 
\begin{equation}
\xi(s, \mu)=\frac{\mathrm{DD}-2 \mathrm{DR}+\mathrm{RR}}{\mathrm{RR}},
\end{equation}
where $\mathrm{DD}$, $\mathrm{DR}$, and $RR$ are the normalized counts of galaxy-galaxy, galaxy-random, and random-random pairs. We further decompose $\xi(s, \mu)$ into multipole moments with 
\begin{equation}
\xi_{\ell}(s)=\frac{2 \ell+1}{2} \int_{-1}^1 \mathrm{~d} \mu \xi(\mathrm{~s}, \mu) \mathrm{P}_{\ell}(\mu)
\end{equation}
where $\mathrm{P}_{\ell}$ is the Legendre polynomial of order $\ell$. The monopole $\xi_0$ and quadrupole $\xi_2$ moments are used in this work.

The calculations described above are carried out with PYCORR\footnote{https://github.com/cosmodesi/pycorr}, which is a Python wrapper for correlation function estimation with \textsc{Corr}\textsc{func} \cite{10.1093/mnras/stz3157} as its two-point counter engines. For the measurement of $\xi(s, \mu)$, we use 241 $\mu$ bins from -1 to 1 and scale-dependent $s$ bins: 4 linearly spaced bins in range $0<s<4~h^{-1}Mpc$, 9 linearly spaced bins in range $4<s<30~h^{-1}Mpc$, and 24 linearly spaced bins in range $30<s<150~h^{-1}Mpc$. The binning scheme is very close to that used in \cite{2024MNRAS.531.3336C,2024MNRAS.531..898P}, which tries to resolve important features in the correlation functions with as few bins as possible. The small number of bins allows for a convergent and reliable estimate of the covariance matrix for the summary statistics based on the N-body simulations.

The 2PCF for an Abacus mock and the CMASS sample are shown in figure~\ref{fig:Emu_vs_cal} and figure~\ref{fig:Emu_vs_CMASS}, respectively.  We will describe the comparison between the measurements and the emulator predictions detailedly in Sec~\ref{sec:emulator_accuracy}.

\subsection{Emulators of the summary statistics}
\label{sec:emulators}

As explained in Sec~\ref{sec:survey_systematics}, all of the galaxy catalogs used in this work should be downsampled to have a constant number density $\bar{n}=2.4\times 10^{-4} h^{3}\rm{Mpc}^{-3}$. However, about $18\%$ of the 53125 HOD variations don't generate galaxy catalogs with a number density higher than $2.4\times 10^{-4} h^{3}\rm{Mpc}^{-3}$, which are discarded because we find our emulator doesn't benefit from them. We note the removal of these models will introduce a non-trivial prior on the HOD parameters, as shown in figure~\ref{fig:barn_hoddist}, we will return to this in section~\ref{sec:priors} and investigate how this non-trial prior impacts our results in Appendix~\ref{sec:pipetest}.

With the summary statistics measured from the Abacus mocks, we can train emulators, which are surrogate models used to efficiently and accurately give predictions for the summary statistics, using flexible machine learning models such as the widely used neural networks \cite{Yuan_2023,2024MNRAS.531.3336C,2024MNRAS.531..898P,Valogiannis_2024} and Gaussian processes \cite{2010ApJ...715..104H,Kwan_2015,2017ApJ...847...50L,2021PhRvD.103l3525R,2022arXiv220712345M,2019ApJ...874...95Z,Zhai_2023,2022MNRAS.515..871Y,2023MNRAS.523.5538Z}. 

We take the galaxy mocks from cosmologies c000, c001, c002, c003, and c004 as test sets, while the remaining mocks of the first and second Latin hypercube samples described in Section~\ref{sec:halo_galaxy_connection} are used as the training and validation sets. That is, there are 2512, 32812, and 8197 mocks for test, training, and validation sets, respectively.
The neural network emulators are trained separately for the MFs and 2PCF, with the cosmological and HOD parameters as the input layer and the concatenated four MFs or monopole and quadrupole of 2PCF as the output layer. Due to the significant variations in the amplitude of the MFs across different orders and in the 2PCF for different separations, it is beneficial to first scale them to a comparable range. This ensures that each data point contributes equally to the loss function, which measures the root mean squared error between each element of the input statistics and the emulator prediction. We standardize both the input parameters and output statistics by subtracting the mean and dividing by the standard deviation. Additionally, we have experimented with scaling the statistics to a range between 0 and 1 and/or using a mean squared error loss function weighted by the variance of the statistics. We find that these alternative approaches do not significantly affect our results.

For both the emulator of MFs and 2PCF, we adopt a neural network of four hidden layers and 512 nodes per layer with a Sigmoid Linear Units activation function. The AdamW optimization algorithm \cite{loshchilov2019decoupledweightdecayregularization}, an improved version of Adam \cite{kingma2017adammethodstochasticoptimization} with properly implemented weight decay regularization, is employed to optimize the network's weights. The neural network is trained with a batch size of 64 and a learning rate of $10^{-3}$. To address learning stagnation, we reduce the learning rate by a factor of 2 every 30 epochs if the validation loss does not improve, continuing this adjustment until the learning rate reaches its minimum value of $10^{-6}$.  The values of these hyperparameters of both the neural network and its training process are chosen based on insights from our previous work \cite{2024MNRAS.531.3336C,Valogiannis_2024}. We also explored alternative hyperparameter choices for the neural network and its training process using Optuna \footnote{\href{https://github.com/optuna/optuna}{https://github.com/optuna/optuna}} to minimize the validation loss. After 200 trials with Optuna, varying the learning rate, number of layers, number of hidden units per layer, dropout rate, and weight decay, we found no improvement over the emulator trained with our default settings.

\subsection{Likelihood}
\label{sec:likelihood}
We assume a Gaussian likelihood for the data vector of the MFs, 2PCF, and their combination. Actually, we have found in Appendix \ref{sec:Gaussian_test} that the likelihood for each statistic can be well approximated with Gaussian. In particular, the log-likelihood is defined as 
\begin{equation}
\label{eq:likelihood}
\mathcal{L}=-\frac{1}{2}\left[\boldsymbol{x}-\boldsymbol{x}(\boldsymbol{\theta})\right] \mathbf{C}^{-1}\left[\boldsymbol{x}-\boldsymbol{x}(\boldsymbol{\theta})\right]^{\top},
\end{equation}
where $\boldsymbol{x}$ can be the data vector measured for the CMASS galaxy sample, the Abacus, or the Uchuu mock galaxy catalog, $\boldsymbol{x}(\boldsymbol{\theta})$ is the output prediction from the emulator for the input parameter set $\boldsymbol{\theta}$, $\mathbf{C}$ is the covariance matrix including three kinds of contributions \cite{2022MNRAS.515..871Y,2024MNRAS.531.3336C,2024MNRAS.531..898P}: 
\begin{equation}
\label{eq:cov}
\mathbf{C} = \mathbf{C_{\rm data}} +  \mathbf{C_{\rm emu}} +  \mathbf{C_{\rm abacus}}.
\end{equation}

\begin{figure}[tbp]
	\centering
	\includegraphics[width=1.0\textwidth]{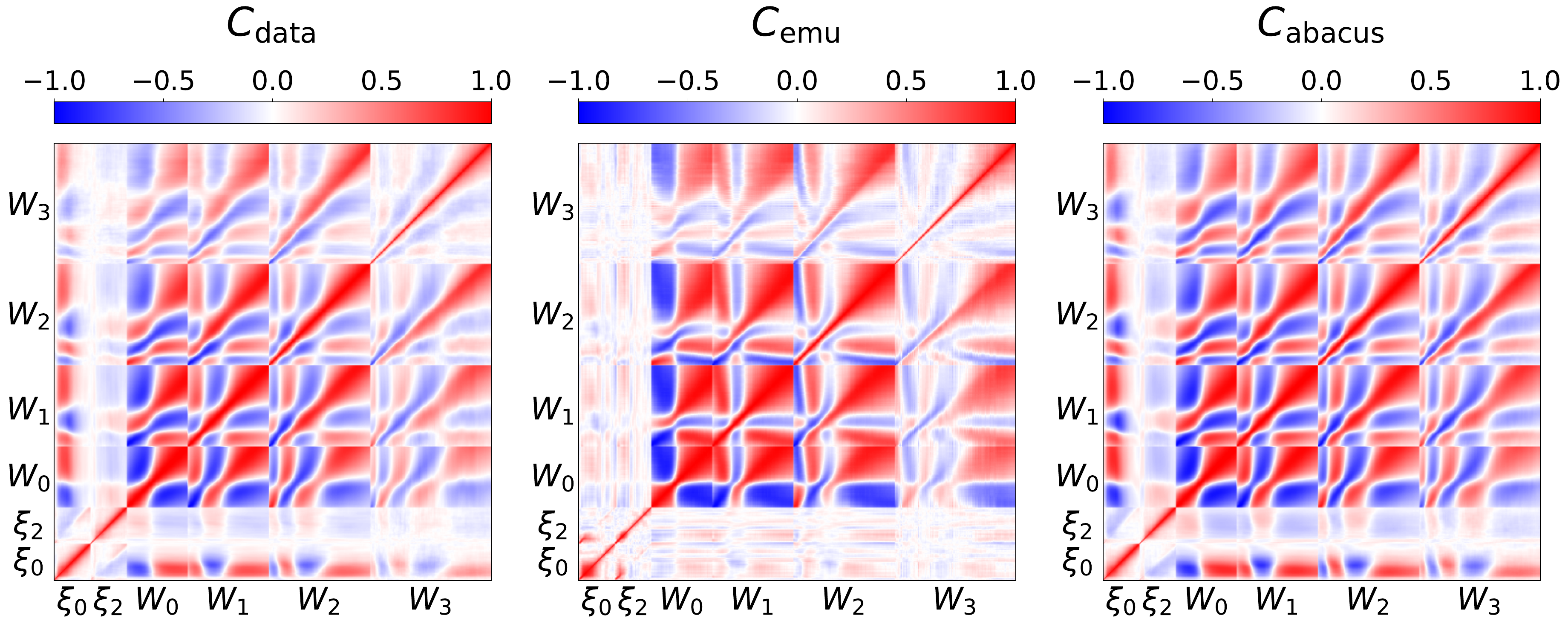}
	\caption{\label{fig:C_data_emu_sim} The corresponding correlation matrices of the three types of covariance matrix described in the text. $C_{\rm data}$ accounts for the sample variance of the CMASS galaxy sample, estimated with the 2048 Patchy mocks. $C_{\rm emu}$ describe the intrinsic error of the emulators, estimated with the 2512 Abacus mocks in the test set. $C_{\rm abacus}$ is the cosmic variance inherited in the emulators because they are trained at a single phase, it is estimated with the 1786 \textsc{Abacus}\textsc{Small} simulations.} 
\end{figure}

The first term $C_{\rm data}$ corresponds to the contribution from sample variance of the CMASS galaxy data, and we estimate this covariance matrix from the 2048 Patchy mocks with
\begin{equation}
\label{eq:c_data}
C_{\mathrm{data}}=\frac{1}{N_{\text {patchy}}-1} \sum_{\mathrm{n}=1}^{N_{\text {patchy }}}\left(\boldsymbol{x}_{\rm patchy}^n-\overline{\boldsymbol{x}}_{\rm patchy}\right)\left(\boldsymbol{x}_{\rm patchy}^n-\overline{\boldsymbol{x}}_{\rm patchy}\right)^{\mathrm{T}},
\end{equation}
where $N_{\rm patchy}=2048$, $\boldsymbol{x}^n_{\rm patchy}$ is the measurement from the $n$th Patch mock, and $\overline{\boldsymbol{x}}_{\rm patchy}$ is the average of statistics. We find the structure of the MFs is very similar to that presented in our previous work, hence we refer the reader to \cite{2022JCAP...07..045L,2023JCAP...09..037L} for a detailed interpretation of the correlations between different density thresholds and between different orders of the MFs. Here we focus on the correlation between the MFs and 2PCF. There exists some correlation between the MFs and $\xi_0$ of intermediate separations and the correlation decreases for very small and large separations. This is because the Gaussian smoothing applied to the galaxy field smears out structure much smaller than the smoothing scale $R_G=15~h^{-1}\rm Mpc$ and enables us to focus on intermediate scales around $15~h^{-1}\rm Mpc$ \footnote{we refer interested readers to figure 2 of \cite{10.1093/mnras/stt2062} for a better understanding of the scale range probed by the MFs with a smoothing scale $R_G$.}. On the other hand, we note that the MFs don't correlate much with $\xi_2$. This is understandable: first, a considerable proportion of the velocity information is smeared out by the isotropic Gaussian smoothing process; second, the integrals~\ref{eq:mfs_def} collect contributions from the infinitesimal volume or area element in an angle-independent way, this means anisotropies introduced by RSD can be buried when adding up contributions that involve the line-of-sight (LoS) direction and those that involve directions orthogonal to the LoS \cite{2013MNRAS.435..531C,2021arXiv210803851J}. We find the two reasons may explain why the MFs only have a weak constraining power on the velocity bias parameters, which is also discussed in Appendix~\ref{sec:HOD_constraints}. This motivates an investigation of the Minkowski tensors, which is a straightforward extension of the MFs \cite{2018ApJ...863..200A,2019ApJ...887..128A}, to extract anisotropies introduced by RSD \cite{2024arXiv241205662L}. We will also explore anisotropic smoothing techniques like what was used in \cite{PhysRevD.109.083535} to leverage symmetries of redshift-space data and recover the lost anisotropic information in future work.

The second term $\mathbf{C_{\rm emu}}$ describes the intrinsic error from the emulator prediction, it is a key ingredient for analysis based on emulators \cite{2019ApJ...874...95Z,Zhai_2023}. Note that here we ignore the dependency of the emulator with the cosmological and HOD parameters and only include a mean correction.  Based on the test mock galaxy catalogs, which consist of a total of 2512 Abacus mocks at cosmologies c000, c001, c002, c003, and c004, we can estimate the emulator covariance with 
\begin{equation}
\label{eq:c_emu}
C_{\mathrm{emu}}=\frac{1}{N_{\text {test}}-1} \sum_{\mathrm{n}=1}^{N_{\text {test }}}\left(\Delta \boldsymbol{x}_{\rm test}^n-\overline{\Delta \boldsymbol{x}}_{\rm test}\right)\left(\Delta \boldsymbol{x}_{\rm test}^n-\overline{\Delta \boldsymbol{x}}_{\rm test}\right)^{\mathrm{T}},
\end{equation}
where $N_{\rm test}=2512$, $\Delta \boldsymbol{x}_{\rm test}^n$ denotes the difference in the data vector between measured from the $n$th Abacus mock and predicted by the emulator, $\overline{\Delta \boldsymbol{x}}_{\rm test}$ is the average over the 2512 test mocks.  

\begin{figure}[tbp]
	\centering
	\includegraphics[width=1.0\textwidth]{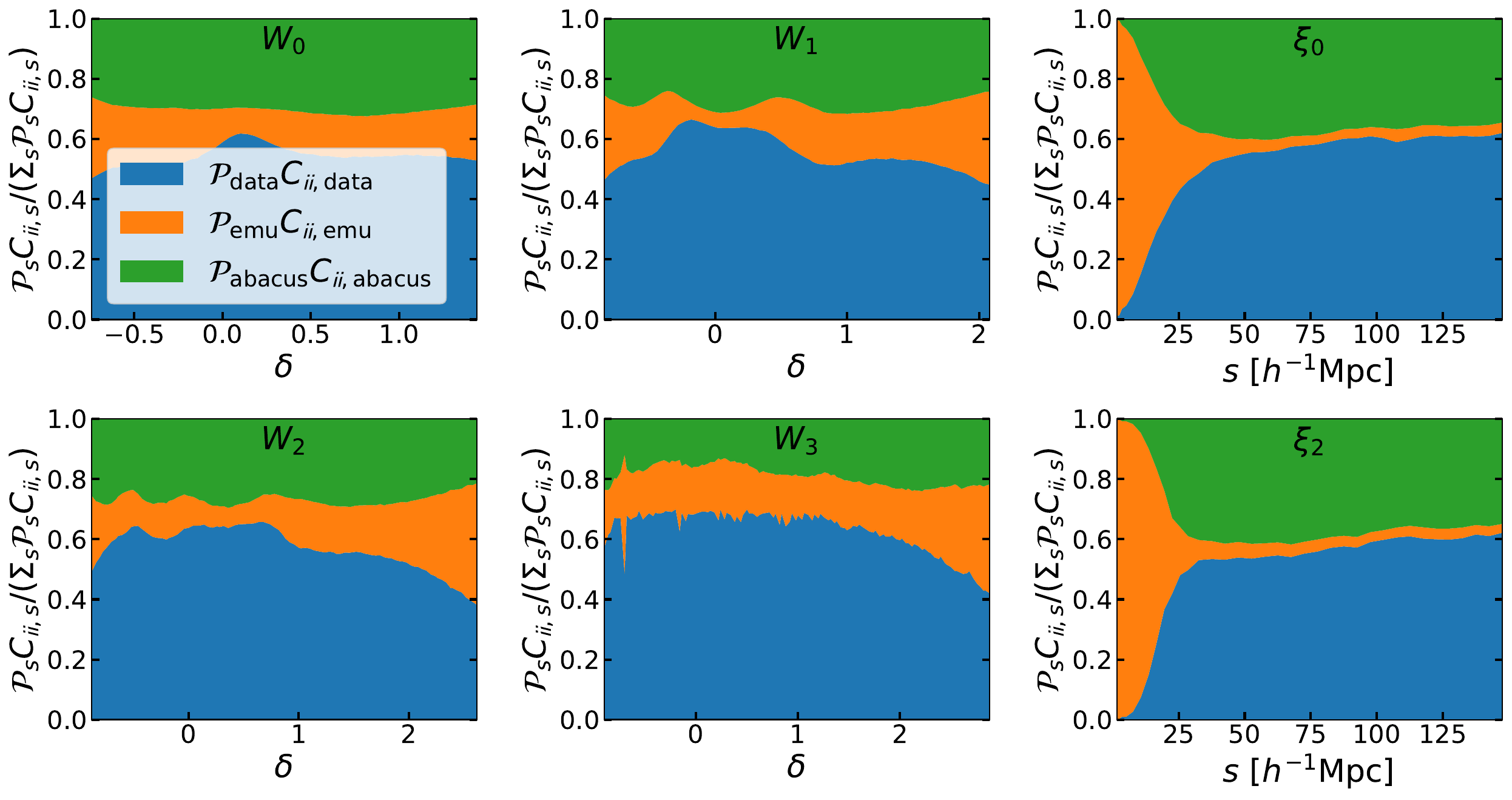}
	\caption{\label{fig:C_frac} Relative contributions from the sample variance of the CMASS data $C_{\rm data}$, the emulator error $C_{\rm emu}$, and the sample variance of the mock catalogs used to train the emulator $C_{\rm abacus}$ for each order of the four MFs (left and middle columns) and each multipole moment of the 2PCF (right column).} 
\end{figure}

The third term is necessary because the emulator is trained based on the simulations at a fixed phase, and there exists a difference in phase between the Abacus mocks and other datasets. This term won't be needed if the emulator is trained using simulations run with different initial conditions so that the uncertainties generated by phase variations can be absorbed into the emulator error \cite{2019ApJ...874...95Z,Zhai_2023}. To estimate this covariance matrix, we first populate dark matter halos of 1786 \textsc{Abacus}\textsc{Small} simulations with HOD galaxies using the best-fit HOD parameter values from \cite{10.1093/mnras/stab3355}. The mock galaxies for each simulation are then downsampled to match the number density $\bar{n}=2.4\times 10^{-4} h^{3}\rm{Mpc}^{-3}$. Finally, we measure the summary statistics for these mock galaxy catalogs and calculate the covariance matrix with 
\begin{equation}
\label{eq:c_abacus}
C_{\mathrm{abacus}}=\frac{1}{N_{\text {abacus}}-1} \sum_{\mathrm{n}=1}^{N_{\text {abacus }}}\left(\boldsymbol{x}_{\rm abacus}^n-\overline{\boldsymbol{x}}_{\rm abacus}\right)\left(\boldsymbol{x}_{\rm abacus}^n-\overline{\boldsymbol{x}}_{\rm abacus}\right)^{\mathrm{T}},
\end{equation}
where $N_{\rm abacus}=1786$, $\boldsymbol{x}^n_{\rm abacus}$ is the measurement from the $n$th \textsc{Abacus}\textsc{Small} simulation, and $\overline{\boldsymbol{x}}_{\rm abacus}$ is the average of all simulations. We note $C_{\mathrm{abacus}}$ is estimated based on mocks in periodic boxes where survey systematics are not included because the CMASS galaxy sample can't be embedded into such a small simulation box.  Despite this, the covariance estimated with these small simulation boxes has successfully reproduced a similar structure to $C_{\rm data}$ as seen in figure~\ref{fig:C_data_emu_sim}. In addition, we find $C_{\rm abacus}$ won't significantly influence our results in Appendix~\ref{sec:pipetest} because the total covariance matrix $C$ is dominated by $C_{\mathrm{data}}$ and $C_{\rm emu}$, this can be seen from figure~\ref{fig:C_frac} and will be discussed later in this section.

The term $C_{\mathrm{abacus}}$ depends on how large the fraction of the simulation box is independently used in our forward model pipeline. As described in Section~\ref{sec:survey_systematics}, we try to increase the usage of the simulation volume by repeatedly rotating the simulation box and cutting the Abacus mock into the geometry of the CMASS North and South samples. In fact, the iteration of the galaxy downsampling process involved in this repeated procedure also helps reduce $C_{\mathrm{abacus}}$ by decreasing the shot noise contribution. To rescale $C_{\mathrm{abacus}}$ estimated from the \textsc{Abacus}\textsc{Small} simulations to the sample variance underlying the emulators, we will utilize the 25 \textsc{Abacus}\textsc{Summit} simulations run at cosmology c000 with different initial conditions. We first assign galaxies to dark matter halos with the same HOD model used for the \textsc{Abacus}\textsc{Small} simulations. Then, the 25 mock galaxy catalogs are all put into the same pipeline used for the Abacus mocks to include survey systematics and measure the summary statistics. Finally, we can rescale $C_{\mathrm{abacus}}$ by
\begin{equation}
\label{eq:rescale_C_abacus}
C^{\prime}_{\mathrm{abacus}}=\frac{\bar{\sigma}^2_{\rm phase}}{\bar{\sigma}^2_{\rm abacus}}C_{\mathrm{abacus}},
\end{equation}
where $\bar{\sigma}^2_{\rm phase}$ is the standard deviations of the statistics from the 25 \textsc{Abacus}\textsc{Summit} simulations averaged over all data bins and $\bar{\sigma}^2_{\rm abacus}$ is the same quantity for the 1786 \textsc{Abacus}\textsc{Small} simulations. Hereafter, we will always use the rescaled variance and still denote it as $C_{\mathrm{abacus}}$, the symbol $C^{\prime}_{\mathrm{abacus}}$ won't be used anymore.

Since the covariance matrices in Eq~\ref{eq:cov} are all estimated with simulations, they are themselves random variables and can lead to a misinterpretation of the derived confidence intervals. To account for this problem, we multiply each of the covariance matrices with a factor $\mathcal{P}$  \cite{10.1093/mnras/stab3540}, 
\begin{equation}
\label{eq:factor_P}
\mathcal{P}=\frac{\left(N_s-1\right)\left[1+B\left(N_{d}-N_\theta\right)\right]}{N_s-N_{d}+N_\theta-1},
\end{equation}
where
\begin{equation}
B=\frac{N_s-N_{d}-2}{\left(N_s-N_{d}-1\right)\left(N_s-N_{d}-4\right)}.
\end{equation}
In the two equations, $N_s$ is the number of mocks used for the estimate of the covariance matrix, $N_{d}$ is the number of data points, and $N_{\theta}$ is the number of parameters to be fit. The final covariance matrix used by us is actually
\begin{equation}
\label{eq:cov_correct}
\mathbf{C} = \mathcal{P}_{\rm data}\mathbf{C_{\rm data}} +  \mathcal{P}_{\rm emu}\mathbf{C_{\rm emu}} +  \mathcal{P}_{\rm abacus}\mathbf{C_{\rm abacus}},
\end{equation}
where $\mathcal{P}_{\rm data}$, $\mathcal{P}_{\rm emu}$, and $\mathcal{P}_{\rm abacus}$ are the corresponding correction factors for each of the three components. Their values are 1.46, 1.34, and 1.55, respectively. We visualize the relative contributions from each of the three terms of Eq~\ref{eq:cov_correct} for each order of the four MFs and each multipole moment of the 2PCF in figure~\ref{fig:C_frac}, where $\mathcal{P}_{s}\mathbf{C}_{ii,s}$ is the diagonal element of each of the three terms on the right-hand side while $\Sigma_s\mathcal{P}_{s}\mathbf{C}_{ii,s}$ is the diagonal element of the covariance matrix on the left-hand side. As seen from this figure, the covariance matrix is dominated by the sample variance of the CMASS data and the emulator error. For the MFs, less than a fraction of $\sim 1/4$ error comes from the sample variance of the Abacus mocks, and the fraction can be as small as $\sim 1/5$ for $W_3$, the last order of the MFs. For the 2PCF, the covariance matrix is significantly dominated by the emulator error for separations smaller than $\sim 10 h^{-1}\rm Mpc$.  

\subsubsection{Priors}
\label{sec:priors}

\begin{figure}[tbp]
	\centering
	\includegraphics[width=1.0\textwidth]{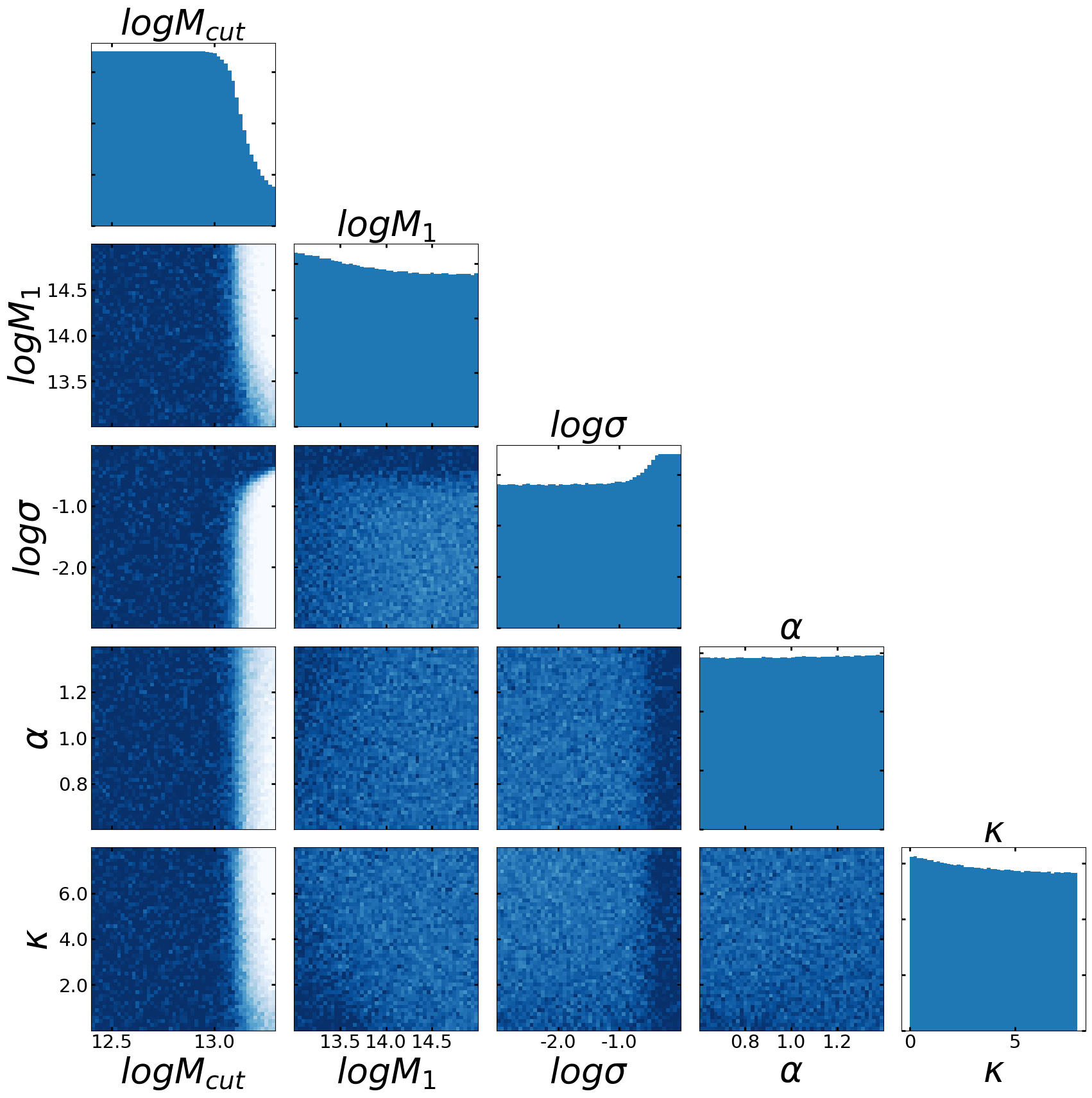}
	\caption{\label{fig:barn_hoddist} The distribution of the five HOD parameters that control the number density of galaxies: $logM_{cut}$, $logM_{1}$, $logM_{\sigma}$, $\alpha$, and $\kappa$. The number density constraint $n(z)\ge 2.4\times 10^{-4}~h^3\rm{Mpc}^{-3}$ poses a non-trivial prior on all of the five parameters. The distribution of $\alpha$ is also non-uniform although this isn't apparent.} 
\end{figure}

We will adopt a Gaussian prior derived from Big Bang Nucleosynthesis (BBN) measurements for $\omega_{\rm b}$ \cite{Aver_2015,Cooke_2018}:
\begin{equation}
\label{eq:PriorOmegab}
\omega_{\mathrm{b}}=0.02268 \pm 0.00038.
\end{equation}
This prior is only used for the analysis of the CMASS data, as the mean value of $\omega_{\rm b}$ will be changed to the true value accordingly when analyzing mocks constructed from simulations. We find in Appendix~\ref{sec:pipetest} whether this prior is used or not only slightly influences the constraints on other cosmological parameters. 

As mentioned in Sec~\ref{sec:simulations}, the dimensionless Hubble parameter is set accordingly so that the CMB acoustic scale $\theta_{*}$
matches the Planck 2018 measurement for all of the cosmologies used to train the emulators. This introduces an implicit prior to the results obtained with the emulators.  The dimensionless Hubble parameter $h$ is thus not a free parameter in our analysis and can be derived from the parameter samples with the $\theta_{*}$ constraint. The constraints on $h$ presented in this work are all derived from the parameter samples using CLASS \cite{Diego_Blas_2011}.

Since only the HOD models with an average number density higher than $n(z)=2.4\times 10^{-4}~h^3\rm{Mpc}^{-3}$ are included for each cosmology when training our emulators, this leads to a non-trivial prior on the HOD parameters and cosmological parameters. To see this, we generate a Latin hypercube with 850000 samples and assign 10000 HOD variations to each of the 85 Abacus cosmologies. The number density of these mock galaxy catalogs is then calculated with the fiducial cosmology, as what has been in Section~\ref{sec:survey_systematics}. After removing models whose number density is smaller than the target value, we show the distribution of the five HOD parameters, which are directly related to the number density of galaxies in figure~\ref{fig:barn_hoddist}. The original distribution of these parameters is supposed to be uniform since they are sampled using the standard Latin hypercube method. However, the number density constraint selectively removes a part of the HOD models, and the distribution is thus no longer uniform.  To account for this, we train an emulator for the number density of galaxies using a similar pipeline to what has been used for the training of the 2PCF and MFs emulators. The selection effect can then be modeled with a prior as a function of the number density 
\begin{equation}
\label{eq:barn_prior}
p(\theta) = (H[\bar{n}(\theta)-0.00024]-1) \times \infty,
\end{equation}
where $H(x)$ is the Heaviside step function and $\bar{n}(\theta)$ is predicted by the number density emulator. This prior function returns $-\infty$ when the galaxy number density of the model specified by $\theta$ is smaller than $0.00024$ and returns $0$ otherwise. Although we find this prior does not significantly influence our constraints on the cosmological parameters in Appendix~\ref{sec:pipetest}, we insist on including it in our analysis for consistency.

We use DYNESTY \cite{10.1093/mnras/staa278} to estimate Bayesian posteriors using the dynamic nested sampling methods \cite{2019MNRAS.483.2044H}, with the likelihood defined as Eq~\ref{eq:likelihood}, the prior function given in Eq~\ref{eq:barn_prior},  a Gaussian prior on $\omega_{\rm b}$ given in Eq~\ref{eq:PriorOmegab}, and flat priors in the range listed in table~\ref{tab:models} on other parameters. For our baseline likelihood analysis, the four extensions of the cosmological parameters $\{\alpha_s,N_{\rm eff},w_0,w_a\}$ are fixed to $\{0.0,3.0146,-1.0,0.0\}$, corresponding to the $\Lambda$CDM limit. Most of the results presented in this work are obtained with this default pipeline, but we also explore other options in this work. The priors used in this work are summarised in Table~\ref{tab:priors}.

\begin{center}
	\small
	\begin{table}[tbp]
	\scalebox{0.9}[1]{
    	\begin{tabular}{|c|c|}
            \hline
            Priors & Descriptions \\
            \hline
              Gaussian prior for $\omega_{\mathrm{b}}$ & $\omega_{\mathrm{b}}=0.02268 \pm 0.00038$ ($\omega_{\mathrm{b}}=0.02230 \pm 0.00038$ for test on Uchuu) \\
              \hline
              Flat prior except $\omega_{\mathrm{b}}$ & Uniform prior distribution with the same bounds listed in Table~\ref{tab:models} \\
              \hline
              Number density constraint & $p(\theta) = (H[\bar{n}(\theta)-0.00024]-1) \times \infty$ \\
            \hline
        \end{tabular}
        }
    \caption{\label{tab:priors} Summary of priors used in this work.}
    \end{table}
\end{center}

\section{Validation of Emulators}
\label{sec:validate_emulator}
Before the application of the emulators to the CMASS data, we will first assess their accuracy and then check whether the true value of cosmological parameters can be recovered for both the internal and external datasets.

\subsection{The accuracy of emulators}
\label{sec:emulator_accuracy}
To visualize the performance of emulators, especially when they are applied to observations, we choose an Abacus mock whose summary statistics are closest to those of the CMASS sample among the test sets at c000 cosmology by minimizing the chi-square value
\begin{equation}
\label{eq:mock_select}
\chi^2=\left[\boldsymbol{x}-\boldsymbol{x}^{\prime}\right] \mathbf{C}^{-1}\left[\boldsymbol{x}-\boldsymbol{x}^{\prime}\right]^{\top},
\end{equation}
where $\boldsymbol{x}$ is the measurement of statistics for the CMASS data, $\mathbf{C}$ is the same covariance matrix as that in Eq~\ref{eq:likelihood}, and $\boldsymbol{x}^{\prime}$ is the data vector measured for the test mocks. In figure~\ref{fig:Emu_vs_cal}, we compare the measured MFs and 2PCF (data points) for this Abacus mock with the predictions (solid lines) from emulators given the true underlying cosmological and HOD parameters. The model predictions show excellent agreement with the data for each of the four MFs at a wide range of density thresholds, as well as for both the monopole and quadrupole of the 2PCF over the full-scale range studied in this work. We further quantify the accuracy of emulators in the lower sub-panels for both the MFs and 2PCF. When evaluated in units of the standard deviation estimated with the 2048 Patchy mocks (data errors, hereafter), the difference in the measurement and prediction is well within $1\sigma$ for most of the data points, with only some bins having $> 1\sigma$ discrepancies. 

We are also interested in the emulator's accuracy when tested with multiple cosmology models and a wide range of HOD models. In particular, we test the emulator with a total of 2512 test mocks at the cosmology c000, c001, c002, c003, and c004. For all of the test mocks, we calculate the differences in the statistics between the measurements and predictions and
plot the $16\%$ to $84\%$ ($1\sigma$ emulator error, dark shaded regions) and $2.5\%$ to $97.5\%$ ($2\sigma$ emulator error, medium shaded regions) quantiles of the differences in the lower sub-panels of fig~\ref{fig:Emu_vs_cal}. Overall, the $1\sigma$ emulator error is much smaller than the $1\sigma$ data error over the full range of density thresholds for each order of the MFs, which is consistent with figure~\ref{fig:C_frac}. We find the $2\sigma$ emulator error is close to the $1\sigma$ data error for most of the threshold range but gradually increases in the rightmost (and/or leftmost) part of the threshold range and can become larger than the $1\sigma$ data error. The emulator error of the 2PCF on small scales is much larger than on large scales. The $1\sigma$ emulator error for $s\lesssim 10 h^{-1}\rm Mpc$ can be larger than the $2\sigma$ data error and dominates the total error budget as seen in figure~\ref{fig:C_frac}.

We note the distribution of the differences between the measurements and predictions is generally symmetric around zero for most of the data bins of both the MFs and 2PCF, and this indicates the emulator does not produce biased predictions statistically.

\subsection{Recovery tests on the Abacus mocks}
\label{sec:recover_abacus}

\begin{figure}[tbp]
	\centering
	\includegraphics[width=1.0\textwidth]{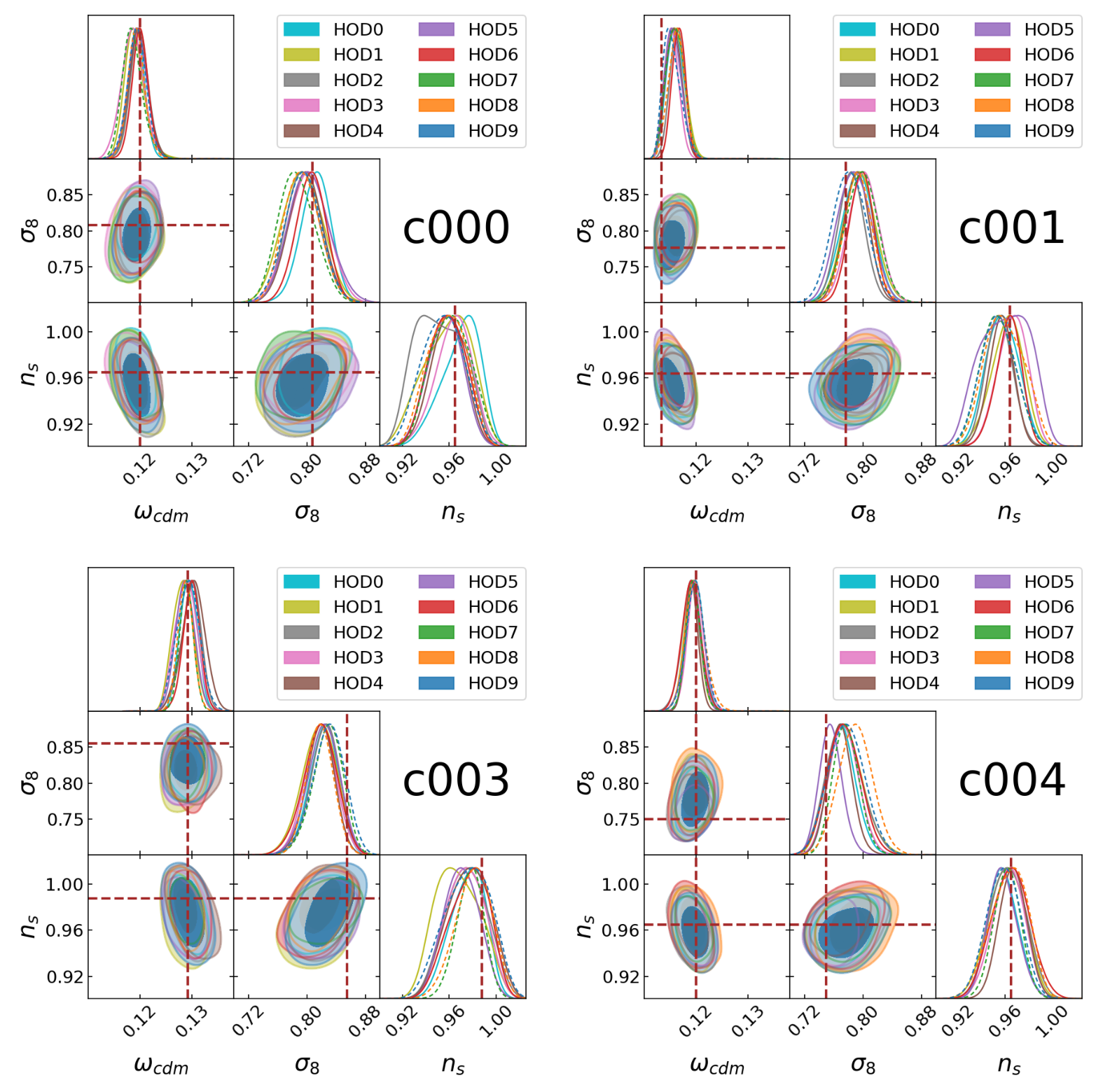}
	\caption{\label{fig:c0_4_recover_test} Marginalized constraints on $\omega_{\rm cdm}$, $\sigma_8$, and $n_s$ from the joint combination of the MFs and 2PCF obtained by applying the emulator to 10 mocks at each of the four test cosmologies: c000 (top left), c001 (top right), c003 (bottom left), and c004 (bottom right). For each test cosmology, we choose the 10 mocks whose HOD parameters are close to the best-fit value from \cite{10.1093/mnras/stab3355}. The vertical and horizontal dashed lines correspond to the true value of the two parameters for each panel. The 2D contours with two different transparencies display the $68\%$ and $95\%$ confidence regions.} 
\end{figure}

In this section, we want to test whether the true cosmological parameters can be inferred with our emulators when applied to the mock galaxy catalogs built from the \textsc{Abacus}\textsc{Summit} simulations. For this purpose, we first select ten mocks from the test set at each of the test cosmologies whose HOD parameter values are close to the best-fit value reported in \cite{10.1093/mnras/stab3355}.  Then, the chosen mocks are fed into our likelihood analysis pipeline, and the mean value of Gaussian prior for $\omega_{\rm b}$ is set to the true value of each test cosmology.

We plot the marginalized constraints on $\omega_{\rm cdm}$, $\sigma_8$, and $n_s$ from the joint combination of the MFs and 2PCF for the chosen 10 mocks at cosmology c000, c001, c003, and c004 in figure~\ref{fig:c0_4_recover_test}. The true underlying parameter values can be recovered within $1~\sigma$ or $2~\sigma$ levels of accuracy for all of these test mocks. The results for c002 are very similar to those for c000 since the two cosmologies share the same value for the three cosmological parameters hence we don't display them here. 

The true value of $\omega_{\rm cdm}$ varies from 0.1134 to 0.1291, while the true value of $\sigma_8$ ranges from 0.7500 to 0.8552 for the four test cosmologies. The variation range of $n_s$ is relatively smaller, from 0.9638 to 0.9876. The success of the parameter recovery over a wide range of cosmologies provides a demonstration of the robustness of our emulator and the likelihood analysis pipeline. We have also done the same recovery test for the MFs and 2PCF emulator individually and found the true parameters can be successfully inferred as well.

\subsection{Recovery tests on the Uchuu mock}
\label{sec:recover_uchuu}
In this work, the emulators for both the 2PCF and MFs are trained with mock galaxy catalogs constructed with the HOD framework. It is thus important to validate that this strong assumption about the galaxy-halo connection does not introduce intractable systematics or bias in the model of the observed data.

We use the mock galaxy catalog from \cite{Zhai_2023}, which is created using the subhalo abundance matching \cite{2004ApJ...609...35K,2004MNRAS.353..189V,2006ApJ...647..201C} (SHAM) method to assign galaxies to dark matter halos of the Uchuu simulations\footnote{\href{http://www.skiesanduniverses.org/Simulations/Uchuu}{http://www.skiesanduniverses.org/Simulations/Uchuu}} \cite{10.1093/mnras/stab1755}. This halo-galaxy connection model is based on the assumed correlation between the stellar mass or luminosity of a galaxy and the properties of the dark matter halo or sub-halo hosting this galaxy. Specifically, galaxies are assigned to dark matter halos using the method of \cite{Lehmann_2016}, where halos are ranked according to a combination of the maximum circular velocity within the halo and the virial velocity. This combination allows for the variation in the amount of assembly bias the galaxies exhibit. 

Uchuu is a suite of cosmological simulations where 2.1 trillion dark matter particles are evolved in a volume of $(2~h^{-1}\rm{Mpc})^3$ with the GreeM N-body code \cite{10.1093/pasj/61.6.1319}, assuming an underlying cosmology described by $\{\omega_{\rm b},\Omega_m,\sigma_8,n_s,h\} = \{0.0486,0.3089,0.8159,0.9667,0.6774\}$. The dark matter halo catalogs are constructed using the Rockstar halo finder \cite{Behroozi_2010}, different from the COMPASO halo finder used for the \textsc{Abacus}\textsc{Summit} simulations. Uchuu shares the same dimensions as those of the \textsc{Abacus}\textsc{Summit} simulation boxes. Thus, the mock galaxies can be directly fed into the same pipeline used for the Abacus mocks.

\begin{figure}[tbp]
	\centering
	\includegraphics[width=1.0\textwidth]{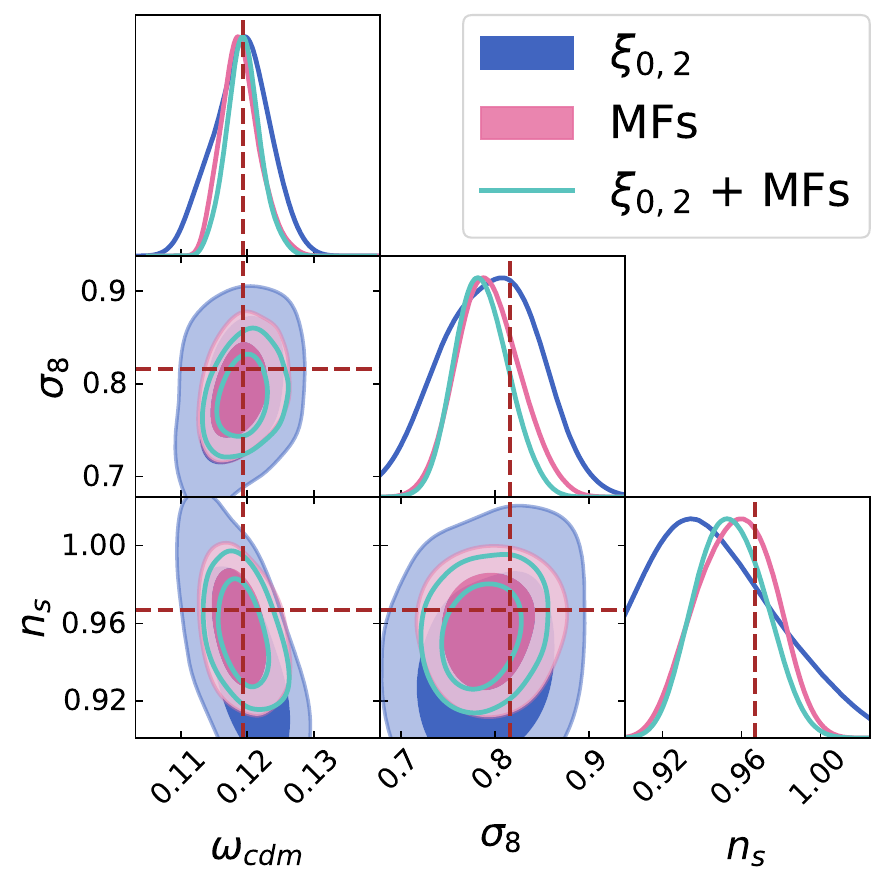}
	\caption{\label{fig:Uchuu_recover} Marginalized posterior derived from the 2PCF (blue), the MFs (pink), and their combination (teal) of the Uchuu mock for the three cosmological parameters: $\omega_{\rm cdm}$, $\sigma_8$, and $n_s$. The vertical and horizontal dashed lines indicate the true value of these parameters. The 2D contours show the $68\%$ and $95\%$ confidence intervals.} 
\end{figure}

For the sampling of the posterior, we also use a Gaussian prior for $\omega_{\rm b}$: $\omega_{\mathrm{b}}=0.02230 \pm 0.00038$, the mean value of $\omega_{\rm b}$ corresponds to the true value of Uchuu simulation and different from what is used for the CMASS sample. We show the marginalized posterior distribution for the Uchuu mock in figure~\ref{fig:Uchuu_recover} obtained with the 2PCF, the MFs, and their combination. The true values of $\omega_{\rm cdm}$, $\sigma_8$, and $n_s$ are correctly inferred for all three cases. We note that the assembly bias is not included in our analysis, however, we find this doesn't lead to any significant bias for the three cosmological parameters. 

This recovery test demonstrates the flexibility of our forward model and its ability to successfully infer the underlying cosmological parameters even though the cosmological simulations are run with a different N-body code, the halo catalogs are built using a different halo finder, and the halo-galaxy connection is modeled through a different framework. We note the recovery test can be successful even without the $\omega_{\rm b}$ prior, which is consistent with our finding that whether the BBN prior is used or not only slightly influences our constraints on $\omega_{\rm cdm}$, $\sigma_8$, and $n_s$ in Appendix~\ref{sec:pipetest}.

\section{Results}
\label{sec:results}
After the validation tests of our pipeline on both internal and external mock data, we apply our emulator model to the CMASS data in this section. 

\subsection{Fits of the summary statistics}
\label{sec:cmass_fits}
\begin{center}
	\small
	\begin{table}[tbp]
	\scalebox{0.88}[1]{
        \begin{tabular}{lccccc}
        \hline Statistic & $\chi^2 /$ dof ($C_{\rm data}$) & $\chi^2 /$ dof ($C_{\rm data} + C_{\rm emu}$) & $\chi^2 /$ dof ($C_{\rm data} + C_{\rm emu}+C_{\rm abacus}$) \\
        \hline MFs (dof=349) & 1.25 & 0.60 & 0.49 \\
        2PCF (dof=61) & 1.53 & 0.70 & 0.52  \\
        MFs + 2PCF (dof=421) & 1.63 & 0.57 & 0.45 \\
        \hline
        \end{tabular}
         }
    \caption{\label{tab:cmass_fits} The goodness-of-fit
parameters (reduced $\chi^2$ values) for the MFs, 2PCF, and their combination, they are calculated by feeding the emulators with the best-fit values from the MFs, 2PCF, and their combination, respectively. ``dof'' denotes the degrees-of-freedom for each statistic. We list values calculated when the covariance matrix only includes the CMASS sample variance (first column), the addition of $C_{\rm data}$ and the emulator error $C_{\rm emu}$ (second column), and the combination of $C_{\rm data}$, $C_{\rm emu}$, and the sample variance of the Abacus mock $C_{\rm abacus}$ (third column).}
    \end{table}
\end{center}

In figure~\ref{fig:Emu_vs_CMASS}, we compare the emulator predictions with the CMASS measurement. Overall, we find the emulator agrees best with the data for the MFs (2PCF) when using the best-fit value from the MFs (2PCF). However, the agreement is weakened when feeding the best-fit value from the MFs (2PCF) to the 2PCF (MFs) emulator. This is understandable since the MFs and 2PCF probe different properties of the LSS. In addition, the MFs focus on structures with scales around the smoothing scale while the 2PCF measures the clustering of galaxies for a wide range of separations. Interestingly, we find large deviations of the MFs emulator from the CMASS data for large thresholds when inputting the best-fit value from the 2PCF. For these high $\delta$ values, the non-Gaussianity becomes important and the 2PCF fails to capture the non-Gaussian information. On the other hand, the Gaussian smoothing used for MFs smears out very small-scale information. That is why the 2PCF emulator can't accurately reproduce both $\xi_0$ and $\xi_2$ of the CMASS data for $s\lesssim10h^{-1}\rm Mpc$ when inputting the best-fit value from the MFs.

Let us focus on the comparison between the emulator prediction and the CMASS measurement for the MFs (2PCF) when using the best-fit value from the MFs (2PCF). The most evident feature is a smooth offset for $W_0$ and $W_1$ between our model and the CMASS data. Take the difference in $W_0$ as an example, if the emulator overpredicts (or underpredicts) the value of $W_0$ at a threshold, it is very likely to overpredict (or underpredict) $W_0$ at the neighboring thresholds as well. Because both the measured and the predicted $W_0$ at the neighboring thresholds are closely correlated with each other, this can be seen from the correlation matrix for $C_{\rm data}$ and $C_{\rm emu}$ shown in figure~\ref{fig:C_data_emu_sim}. Most of the data bins for $W_2$ and $W_3$ are reasonably fitted, with fluctuations centered at the measurement.  For the 2PCF, the emulator has a clear underprediction of the monopole for $s\gtrsim 80 h^{-1}\rm{Mpc}$. The same mismatch can be found between the average of the Patchy mocks and the CMASS sample, which was previously found in \cite{2016MNRAS.456.4156K,10.1093/mnras/stw2372}. Similarly, \cite{2024MNRAS.531..898P} found their 2PCF emulator trained with the \textsc{Abacus}\textsc{Summit} simulations underpredicted the observed clustering in the same scale range. In addition, the model of the 2PCF based on Convolutional Lagrangian Perturbation Theory underpredicted the large-scale clustering of the CMASS data as well~\cite{10.1093/mnras/stx883}. While some of these offsets could result from statistical fluctuations, the possible existence of residual systematic effects~\cite{2019arXiv190906396L}, which are not corrected with the weights assigned to each galaxy, can also produce signals not modeled by our emulator.

We report the goodness-of-fit parameters in table~\ref{tab:cmass_fits} for the best fits to the CMASS data. When the covariance matrix only includes the contribution from the CMASS sample variance ($C_{\rm data}$), we obtained values 1.53, 1.25, and 1.63 for the 2PCF, MFs, and their combination. The value for the 2PCF is larger than those reported in \cite{2022MNRAS.515..871Y,2024MNRAS.531..898P,Valogiannis_2024}, where $\chi^2/dof$ was found to be in the range of 1.06 to 1.15. Nevertheless, we believe our value is reasonable considering the significant tension between the BOSS 2PCF monopole and our best fit on large scales seen in figure~\ref{fig:Emu_vs_CMASS}. The value for the MFs is smaller than that for the density split clustering ($\chi^2/dof=1.67$ \cite{2024MNRAS.531..898P}) and also for the wavelet scattering transform ($\chi^2/dof=1.37$ \cite{Valogiannis_2024}). However, the goodness-of-fit values for the summary statistics and their combination are all smaller than one once including the error budget from the emulator error ($C_{\rm data} + C_{\rm emu}$). This indicates an overestimate of the emulator error, which may be more properly estimated with the Abacus mocks whose summary statistics are close to those of the CMASS data. A similar approach was adopted in \cite{2022MNRAS.515..871Y}, where the error is estimated using only HOD catalogs whose likelihood is high with respect to the CMASS measurement. Our estimate of $C_{\rm emu}$ is more conservative because we have included the error around regions near the edge of the priors, where the emulator is most likely to fail to correctly study the statistics. The goodness-of-fit values are further reduced after the addition of the sample variance for the Abacus mocks ($C_{\rm data} + C_{\rm emu} + C_{\rm abacus}$), which is a suggestion that the double counting of sample variance in the CMASS data and the Abacus mocks may lead to an overestimate of error in the analysis. We will investigate this issue in future work, and we note the phase correction routine used by \cite{2022MNRAS.515..871Y,Valogiannis_2024} may provide a method to alleviate this problem.

\subsection{Constraints on $\Lambda$CDM}
\label{sec:base_results}
\begin{center}
	\small
	\begin{table}[tbp]
	\scalebox{0.88}[1]{
    	\begin{tabular}{ccccccc}
            \hline & \multicolumn{2}{c}{ MFs } & \multicolumn{2}{c}{ 2PCF } & \multicolumn{2}{c}{ 2PCF + MFs } \\
            \hline Parameter & best-fit & mean $\pm \sigma$ & best-fit & mean $\pm \sigma$ & best-fit & mean $\pm \sigma$ \\
            $\omega_{\mathrm{b}}$ & 0.0221 & $0.02268\pm 0.00036$ & 0.0226 & $0.02270\pm 0.00039$ & 0.0228 & $0.02273\pm 0.00035 $ \\
            $\omega_{\mathrm{cdm}}$ & 0.1166 & $0.1174^{+0.0020}_{-0.0023}$ & 0.1186 & $0.1166^{+0.0040}_{-0.0045} $ & 0.1156 & $0.1172^{+0.0020}_{-0.0023}$ \\
            \\[-1em]
            $\sigma_8$ & 0.8044 & $0.799\pm 0.029$ & 0.8008 & $0.773^{+0.047}_{-0.052}$ & 0.7908 & $0.783\pm 0.026$ \\
            \\[-1em]
            $n_s$ & 0.9709 & $0.963^{+0.024}_{-0.016}$ & 0.9688 & $0.951^{+0.023}_{-0.033}$ & 0.9744 & $0.966^{+0.019}_{-0.015}$ \\
            \\[-1em]
            \hline
            \\[-1em]
            $h$ & 0.6898 & $0.6882^{+0.0075}_{-0.0065}$ & 0.6838 & $0.691\pm 0.014$ & 0.6944 & $0.6889^{+0.0075}_{-0.0065}$ \\
            \\[-1em]
            $f \sigma_8$ & 0.464 & $0.462\pm 0.018$ & 0.465 & $0.446\pm 0.031$ & 0.455 & $0.453\pm 0.016$ \\
            \hline
        \end{tabular}
        }
    \caption{\label{tab:base_constraints} The best-fit (maximum a posteriori), mean, and $68\%$ confidence interval values of the four basic cosmological parameters $\omega_{\rm b},\omega_{\rm cdm},\sigma_8,n_s$ and two derived cosmological parameters $h$ and $f\sigma_8$ for the MFs, 2PCF, and their combination, marginalized over all HOD parameters.}
    \end{table}
\end{center}

\begin{figure}[tbp]
	\centering
	\includegraphics[width=1.0\textwidth]{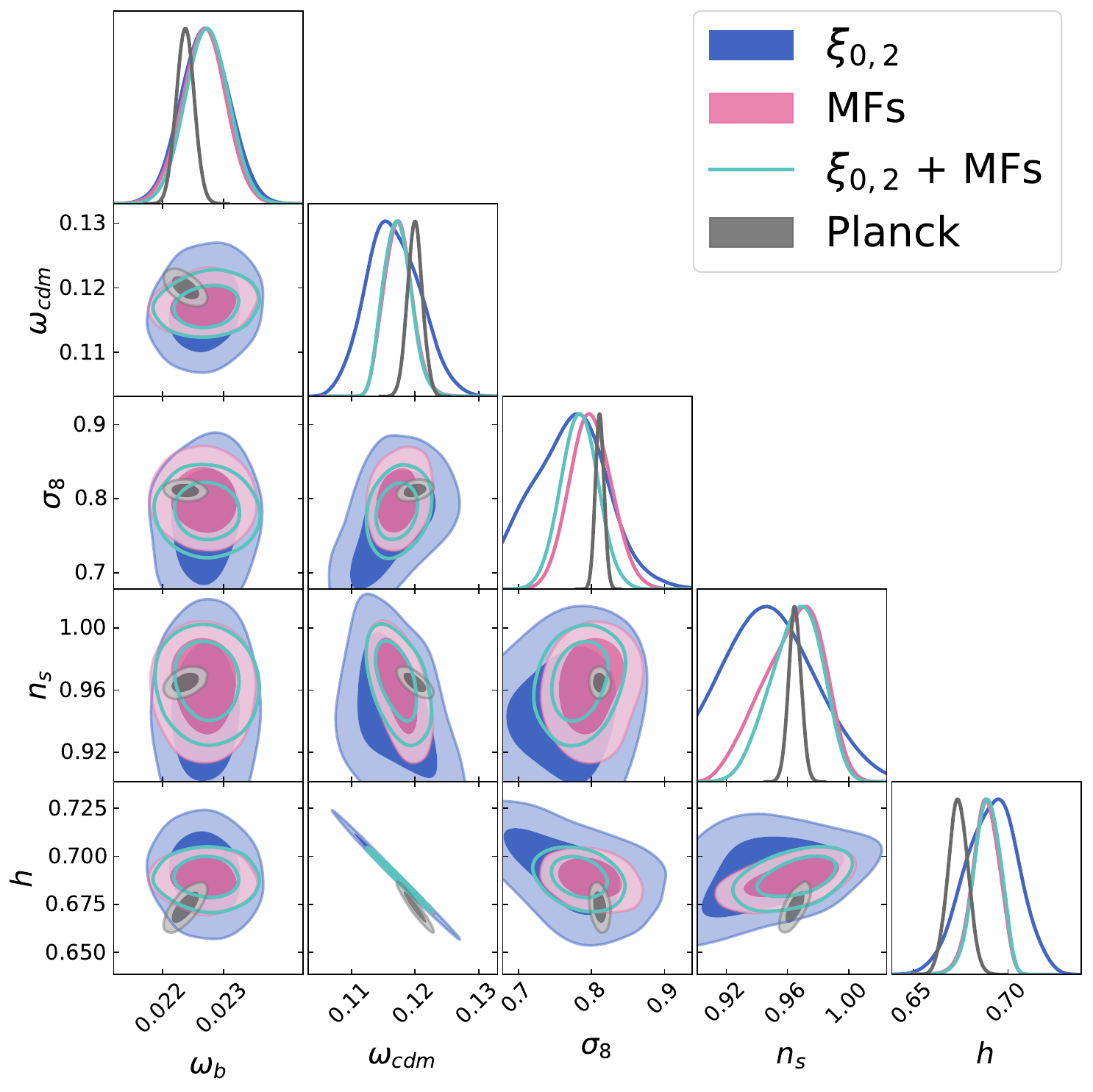}
	\caption{\label{fig:CMASS_fit} Marginalized constraints from the 2PCF (blue), the MFs (pink), their combination (teal) of the BOSS CMASS galaxy sample, and the PLANCK$\_$TTTEEE$\_$LOWL$\_$LOWE likelihood from Planck 2018 (grey). A BBN Gaussian prior $\mathcal{N}\left(0.02268, 0.00038\right)$ is used for $\omega_{\rm b}$ in our analysis of the CMASS data. The 2D contours correspond to the $68\%$ and $95\%$ confidence intervals.} 
\end{figure}

\begin{figure}[tbp]
	\centering
	\includegraphics[width=1.0\textwidth]{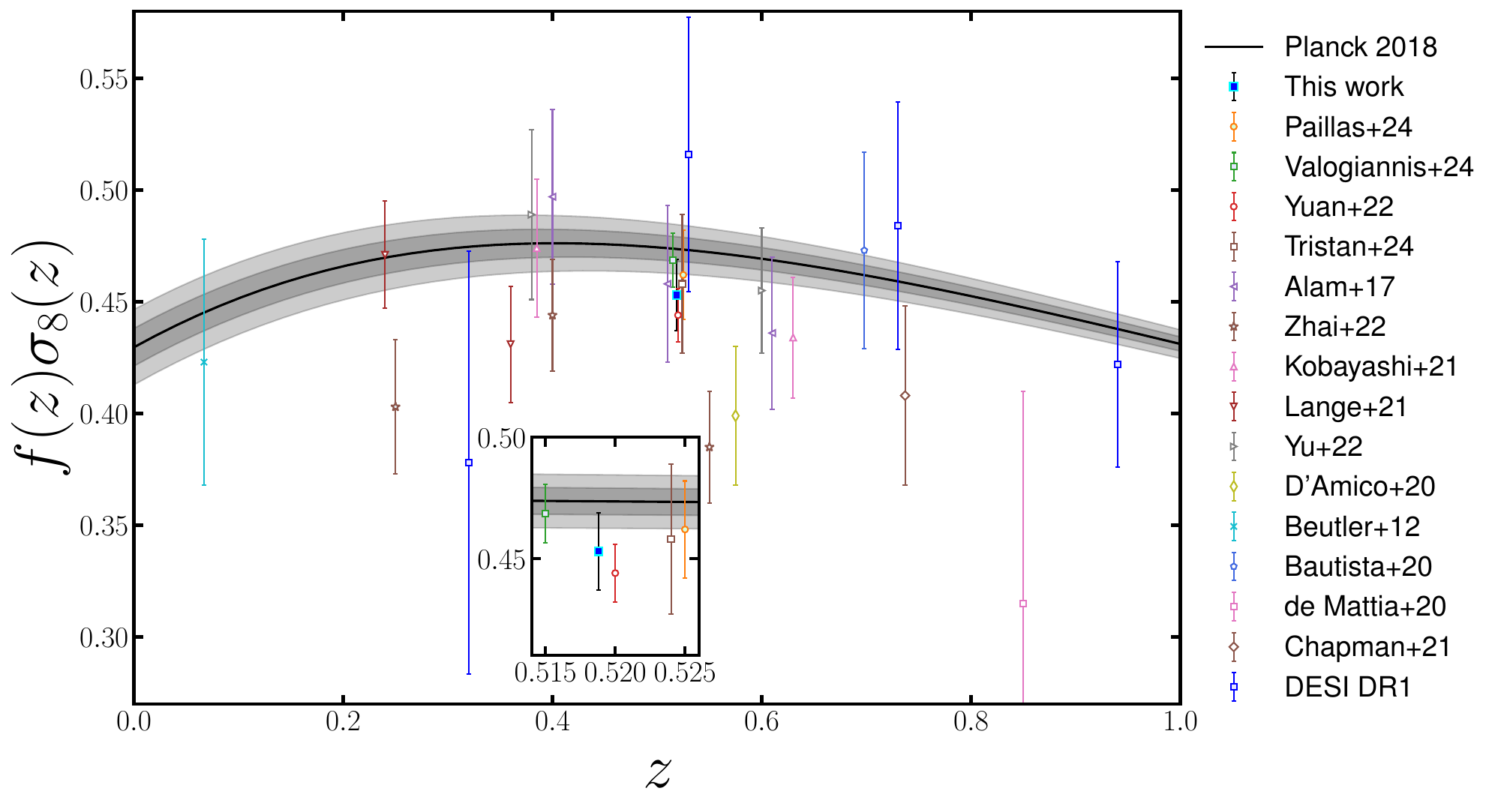}
	\caption{\label{fig:fsigma8} The marginalized posterior means and errors of the structure growth rate $f(z)\sigma_8(z)$ from the combination of the 2PCF and MFs for the BOSS CMASS galaxy sample, together with those from the four recent analyses of CMASS sample based on the  \textsc{Abacus}\textsc{Summit} simulations: the density split clustering statistic (Paillas+24 \cite{2024MNRAS.531.3336C,2024MNRAS.531..898P}), the wavelet scattering transform statistic (Valogiannis+24 \cite{Valogiannis_2024}), the small-scale 2PCF (Yuan+22 \cite{2022MNRAS.515..871Y}), and the void-galaxy cross-correlation function (Tristan+24 \cite{2024arXiv240703221F}). The comparison between this work and these Abacus-based works is better visualized in the inset. In addition, results from the consensus analysis (Alam+17 \cite{10.1093/mnras/stx721}) and other analyses of the BOSS data using simulation-based methods (Zhai+22 \cite{Zhai_2023}, Kobayashi+21 \cite{PhysRevD.105.083517}, Lange+21 \cite{Lange2021}), the halo perturbation theory model (Yu+22 \cite{Yu_2023}), and the Effective Field Theory model (D'Amico+20 \cite{dAmico_2020}). Results from other data sets are also included: the 6dF Galaxy Survey (Beutler+12 \cite{10.1111/j.1365-2966.2012.21136.x}),  the completed SDSS IV extended Baryon Oscillation Spectroscopic Survey (Bautista+20 \cite{10.1093/mnras/staa2800}, de Mattia+20 \cite{10.1093/mnras/staa3891}, Chapman+21 \cite{10.1093/mnras/stac1923}), the Dark Energy Spectroscopic Instrument (DESI Data Release 1) \cite{2024arXiv241112021D}, and the Planck CMB measurements (Greybands show the $68\%$ and $95\%$ confidence ranges) \cite{2020A&A...641A...6P}. The redshifts of some overlapped data points are slightly shifted for better visualization.} 
\end{figure}

In this section,  we present the results obtained using the baseline pipeline setting for the likelihood analysis described in Section~\ref{sec:priors}. To discriminate from those presented in the following Section~\ref{sec:cos_extend}, we emphasize that the values of $\alpha_s,N_{\rm eff},w_0$, and $w_a$ are fixed to 0.0, 3.0146, -1.0, and 0.0, respectively. We infer the four cosmological parameters $\omega_{\rm b}$, $\omega_{\rm cdm}$, $\sigma_8$, $n_s$ and the seven HOD parameters at the same time. We describe the constraints on the cosmological parameters here and leave the discussion on results for the HOD parameters in Appendix~\ref{sec:HOD_constraints}. The 2D marginalized posterior probability distributions for the four cosmological parameters are displayed in figure~\ref{fig:CMASS_fit}, together with that derived for the Hubble parameter using the $\theta_{*}$ constraint described in Section~\ref{sec:priors}. They are obtained from the 2PCF, MFs, and their combination. To facilitate comparison, we also plot the base-CDM constraints from the PLANCK$\_$TTTEEE$\_$LOWL$\_$LOWE likelihood (Planck 2018). We list the corresponding best-fit values, mean values, and 68$\%$ confidence intervals for these parameters in table~\ref{tab:base_constraints}. 

Overall, we find our constraints are consistent with Planck 2018 results within $\sim 1\sigma$ level for the 2PCF, MFs, and their combination. The constraints from the 2PCF and MFs are also consistent with each other. The constraint on $\omega_{\rm b}$ is dominated by the Gaussian prior. The strong degeneracy between $\omega_{\rm cdm}$ and $h$ exists for both the MFs and 2PCF, which can also be seen for the density-split clustering \cite{2024MNRAS.531..898P} and the wavelet scattering transform \cite{Valogiannis_2024}. This strong degeneracy is related to the dependence of the Hubble parameter on $\omega_{\rm cdm}$ when the $\theta_{*}$ constraint is imposed at a standard $\Lambda$CDM cosmology \cite{2024MNRAS.531..898P}.

Compared with the 2PCF, the MFs provide a factor of 2.0, 1.7, 1.4, and 2.0 times tighter constraints on $\omega_{\rm cdm}$, $\sigma_8$, $n_s$, and $h$, respectively. This is consistent with \cite{2025MNRAS.537.3553A}, where they found the MFs of weak lensing convergence maps also have stronger constraining power on $\Omega_m$ and $S_8=\sigma_8\sqrt{\Omega_m/0.3}$ than two-point statistics. The amplitudes of the MFs are sensitive to the shape of the linear matter power spectrum and thus to $\omega_{\rm cdm}$ and $n_s$ \cite{2003ApJ...584....1M,Appleby_2020,2022ApJ...928..108A}. In redshift space, the amplitudes are also sensitive to the growth rate of the structure (thus $\omega_{\rm cdm}$) and galaxy bias (HOD parameters) \cite{1996ApJ...457...13M,2013MNRAS.435..531C}. They are practically insensitive to the amplitude of the power spectrum, that is, $\sigma_8$. However, the shapes of the MFs are very sensitive to $\sigma_8$ because it controls the amplitude of the departure from the Gaussian shape of the MFs and determines the position and scaling of the shape through the threshold parameter $\delta$ \cite{2013MNRAS.435..531C,2023JCAP...09..037L}. Extra constraining power on $\omega_{\rm cdm}$ and $n_s$ comes from the non-Gaussian information extracted by the MFs\footnote{The non-Gaussian information is not necessarily extracted only from the departure from the Gaussian shape of the MFs, because the amplitude of the MFs also receives contributions from the trispectrum \cite{2003ApJ...584....1M,2013MNRAS.435..531C,PhysRevD.104.103522}.}, as found in \cite{2022JCAP...07..045L,2023JCAP...09..037L}. Otherwise, the MFs are anticipated to provide similar constraints (theoretically) on the two parameters to those from the 2PCF. The strong constraint on $h$ may stem from the $\theta_{*}$ prior because we do not expect to capture more accurate information of $h$ than the 2PCF with a single smoothing scale, which is similar to what was found in \cite{Valogiannis_2024}.

When combining the 2PCF and MFs, we find the MFs dominate the constraints on $\omega_{\rm cdm}$, $n_s$, and $h$. The information provided by the 2PCF only slightly tightens constraints on these parameters, which indicates this part of the information embedded in the MFs is highly overlapped with that from the 2PCF. However, we note the MFs can provide more complementary information for the 2PCF when the number density of the galaxy sample is higher, and the smoothing scale is smaller in another ongoing work of ours. As mentioned above, the amplitudes of the MFs are insensitive to $\sigma_8$, that is, the Gaussian information extracted by the MFs has weak constraining power on $\sigma_8$. This part of the information is supplemented by the 2PCF, and we find the 2PCF indeed helps tighten the constraint on $\sigma_8$. Compared with the 2PCF, the combination of the two statistics improves the constraints on $\omega_{\rm cdm}$, $\sigma_8$, $n_s$, and $h$ by a factor of 2.0, 1.9, 1.6, and 2.0.

To make a broader comparison with existing constraints from previous studies in the literature, we also derive the parameter combination $f\sigma_8$ from the samples in our chains, where $f$ is the linear growth rate. From the combination of the MFs and 2PCF, we get a $4\%$ constraint:
\begin{equation}
f \sigma_8\left(z_{\mathrm{eff}}=0.519\right)=0.453 \pm 0.016.
\end{equation}
We compare our constraint on $f\sigma_8$ with the other four analyses of the CMASS sample based on the \textsc{Abacus}\textsc{Summit} simulation in table~\ref{tab:constraints_comparison}. We also plot our result together with the Planck 2018 result \cite{2020A&A...641A...6P} and a series of previous analyses of LSS observations in figure~\ref{fig:fsigma8}. Although the consistency between our result and the Planck 2018 result is slightly larger than $1\sigma$ level, we find the inferred $f\sigma_8$ agrees with that from the other four Abacus-based analyses within a $1\sigma$ level\footnote{Since the values of $z_{\rm eff}$ are so close that we can directly compare the mean values of $f\sigma_8$.}. It is also in agreement with the latest measurement from the Dark Energy Spectroscopic Instrument (DESI Data Release 1) \cite{2024arXiv241112021D}.

Focusing on the comparison with the consensus BOSS clustering analysis \cite{10.1093/mnras/stx721},  our constraint is consistent with and $\sim 2.4$ times tighter than their result at $z_{\rm eff}=0.51$, which is obtained using BOSS large-scale RSD + BAO. In figure~\ref{fig:fsigma8}, we also show results from the other three simulation-based analyses of the BOSS data, these small-scale studies utilize methodologies including the correlation function emulator \cite{Zhai_2023} trained with the AEMULUS simulation suite \cite{DeRose_2019}, the full-shape galaxy power spectrum emulator \cite{PhysRevD.105.083517} built on the Dark Quest simulation suite \cite{Nishimichi_2019}, and the evidence modeling approach \cite{Lange2021} also based on the AEMULUS simulation suite. In addition, we display the large-scale analyses based on the power spectrum multipoles modeled using Eulerian perturbation theory \cite{Yu_2023} and the Effective Field Theory of Large-Scale Structures \cite{dAmico_2020}. Although it is difficult to make a straightforward comparison with these studies due to the large differences existing in the adopted methodology and analysis choices, we notice that our analysis is one of the most stringent constraints. On the other hand, we note most of the analyses of BOSS data, including ours, tend to underpredict the growth rate relative to the Planck 2018 result, although mainly at a low statistical significance except for the measurement from \cite{dAmico_2020} and the highest redshift bin of \cite{Zhai_2023}.

\subsection{Constraints on $\Lambda$CDM extensions}
\label{sec:cos_extend}
\begin{figure}[tbp]
	\centering
	\includegraphics[width=1.0\textwidth]{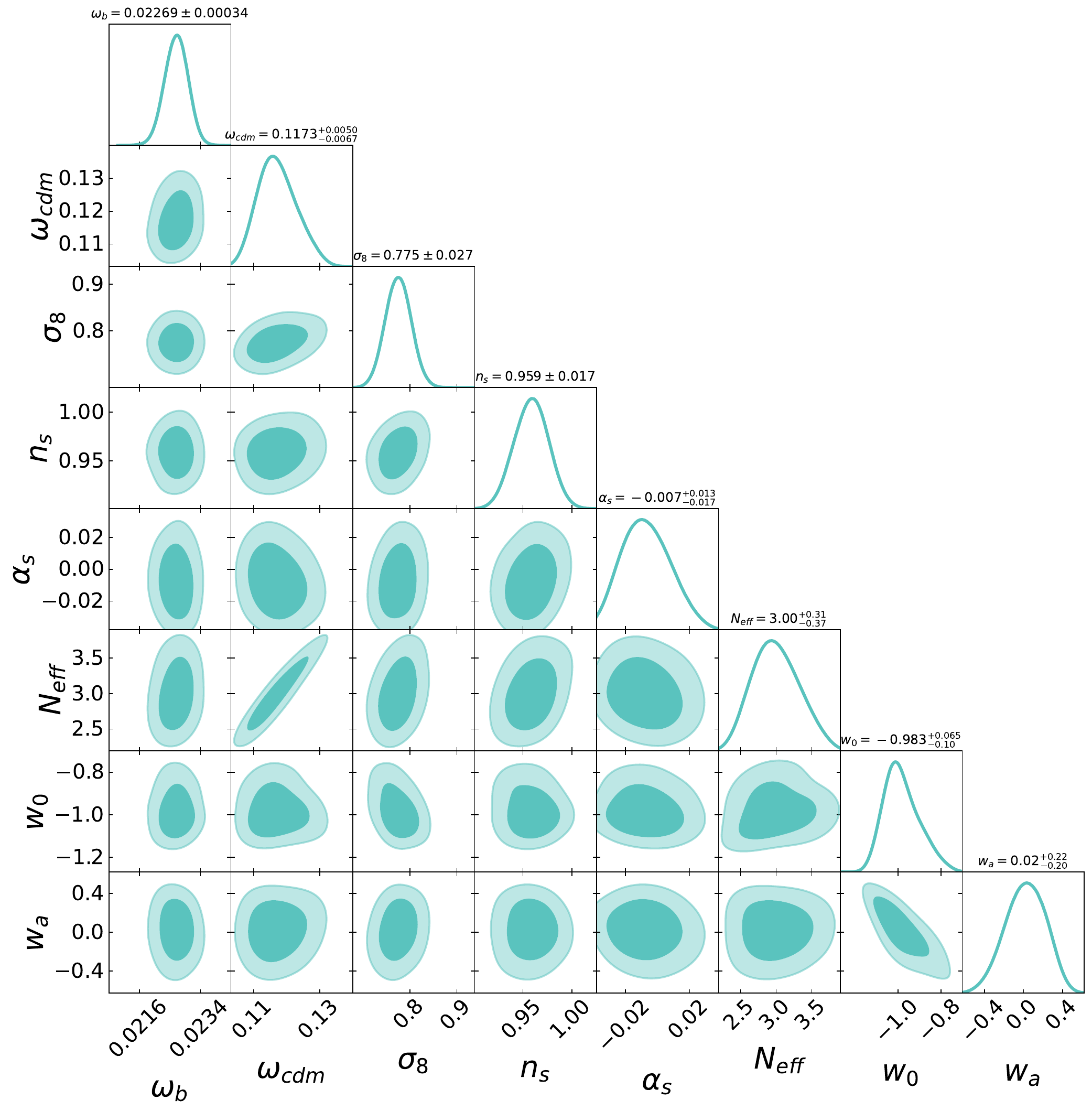}
	\caption{\label{fig:CMASS_fit_extend_cos} Marginalized constraints from the combination of the MFs and 2PCF for the BOSS CMASS galaxy sample. A BBN Gaussian prior $\mathcal{N}\left(0.02268, 0.00038\right)$ is used for $\omega_{\rm b}$. The 2D contours correspond to the $68\%$ and $95\%$ confidence intervals.} 
\end{figure}

Since the \textsc{Abacus}\textsc{Summit} simulations were run for various sets of cosmologies in an 8-dimensional parameter space, including the four extensions to the standard $\Lambda$CDM model: $\alpha_s,N_{\rm eff},w_0$, and $w_a$. In contrast to the baseline setting used in section~\ref{sec:base_results}, we here allow the four extended cosmological parameters to vary with the other four basic cosmological parameters plus the seven HOD parameters at the same time. The marginalized posterior distributions obtained from the combination of the MFs and 2PCF of the CMASS data are displayed in figure~\ref{fig:CMASS_fit_extend_cos}.

Compared with results presented in section~\ref{sec:base_results}, we find the constraints on $\omega_{\rm cdm}$, $\sigma_8$, and $n_s$ are consistent with each other, no matter whether the extended cosmological parameters are fixed or not. The weakened constraints on the three parameters are expected due to the extended parameter space and the introduction of more parameter degeneracies. The inferred constraints: $\alpha_s=-0.007^{+0.013}_{-0.017}$, $N_{\rm eff}=3.00^{+0.31}_{-0.37}$, $w_0=-0.98^{+0.07}_{-0.10}$, and $w_a=0.02^{+0.22}_{-0.20}$ are all within $1\sigma$ agreement with the Planck 2018 measurement and the fiducial $\Lambda$CDM cosmology used in this work. 

\section{Discussions}
\label{sec:discussions}
In this section, we compare our results with other Abacus-based analyses, previous analysis of the MFs for the CMASS data, and the Fisher matrix forecasts for the MFs.
\subsection{Comparison with other Abacus-based analyses}
\begin{center}
	\small
	\begin{table}[tbp]
	\scalebox{0.84}[1]{
    	\begin{tabular}{cccccc}
            \hline 
            Statistics &  MFs + 2PCF & density-split + 2PCF & WST + 2PCF & small-scale 2PCF & VG-CCF\\
            $\omega_{\mathrm{cdm}}$ & $0.1172^{+0.0020}_{-0.0023}$ & $0.1201\pm 0.0022$ & $0.1241 \pm 0.0011$ & $0.115\pm0.004$ & $0.1235\pm0.0037$\\
            $\sigma_8$ & $0.783\pm 0.026$ & $0.792\pm 0.034$ & $0.795 \pm 0.019$ & $0.756\pm0.016$ & $0.777^{+0.047}_{-0.062}$\\
            $n_s$ & $0.966^{+0.019}_{-0.015}$ & $0.970\pm0.018$ & $0.924\pm0.01$ & $0.94\pm0.02$ & -\\
            \\[-1em]
            \hline
            \\[-1em]
            $h$ & $0.6889^{+0.0075}_{-0.0065}$ & $0.6793 \pm 0.0070$ & $0.669\pm0.0049$ & - & -\\
            \\[-1em]
            $f \sigma_8$ & $0.453\pm 0.016$ & $0.462\pm 0.020$ & $0.469\pm0.012$ & $0.444\pm0.012$ & $0.458^{+0.029}_{-0.033}$\\
            & $(z_{\rm eff}=0.519)$ & $(z_{\rm eff}=0.525)$ & $(z_{\rm eff}=0.515)$ & $(z_{\rm eff}=0.52)$ & $(z_{\rm eff}=0.525)$\\
            \hline
        \end{tabular}
        }
    \caption{\label{tab:constraints_comparison} Comparison of the mean and $68\%$ confidence interval values between the combination of 2PCF with the MFs (first column), the density-split clustering \cite{2024MNRAS.531..898P} (second column), the wavelet scattering transform \cite{Valogiannis_2024} (third column), as well as the sole small-scale 2PCF \cite{2022MNRAS.515..871Y} and the void-galaxy cross-correlation function \cite{2024arXiv240703221F}.}
    \end{table}
\end{center}

In table~\ref{tab:constraints_comparison}, we compare the mean and $68\%$ confidence interval values from the combination of 2PCF and the MFs and results from previous analysis of the CMASS galaxy sample based on emulators trained using the \textsc{Abacus}\textsc{Summit} simulations (Abacus-based works, hereafter). We see the best agreement between our results and those from the combination of density-split clustering and the 2PCF (density-split + 2PCF)\cite{2024MNRAS.531..898P}, with the differences in the mean values all smaller than $1~\sigma$. Our constraints on all five parameters are tighter than or comparable to the density-split + 2PCF. We note this may not be a fair comparison because the environment-based assembly bias parameters are implemented in their analysis and allowed to vary during the fit. The added degeneracies between the assembly bias parameters and other parameters can weaken the constraining power of the density-split + 2PCF.

Compared with results from the combination of the wavelet scattering transform (WST) and the 2PCF \cite{Valogiannis_2024}, we find tensions in the mean values of $\omega_{\rm cdm}$, $n_s$, and $h$. There exists $2.9~\sigma$, $2.1~\sigma$, and $2.3~\sigma$ level of differences in the mean values of $\omega_{\rm cdm}$, $n_s$, and $h$ between our and their results. From the 2D marginalized posterior distribution shown in Fig. 8 of their paper \cite{Valogiannis_2024}, we find the mean value of $\omega_{\rm cdm}$ and $h$ is mainly determined by the information from the WST while the mean value of $n_s$ is dominated by the 2PCF. Although the mean values of the three parameters are mainly nailed down by the MFs in this work, we do find the 2PCF prefers a lower value of $n_s$.  We can't give a straightforward explanation about the origin of these tensions in parameters; it will be investigated in our ongoing work. Despite these tensions, the combination of WST and 2PCF gives tighter constraints on all of these parameters than our results. This may be due to the smaller emulator error in their emulators since a much larger number of galaxy mocks are used to train their emulators. Another reason for this is that a smaller Gaussian-like smoothing scale $\sim 8h^{-1}\rm Mpc$ was adopted in their work, which may allow them to extract more information. It is also possible that there is more information embedded in the combination of the WST and 2PCF than the MFs and 2PCF.

We also see reasonable agreement between our results and those from the small-scale 2PCF \cite{2022MNRAS.515..871Y} and the void-galaxy cross-correlation function \cite{2024arXiv240703221F}. The differences in the mean values are all within $2\sigma$. The tighter constraints on $\sigma_8$ and $f\sigma_8$ from the small-scale 2PCF result from the hybrid MCMC $+$ emulator framework developed in \cite{2022MNRAS.515..871Y}, which can decrease the emulator error by constructing the emulator within the high-likelihood region of cosmology and HOD parameter space.

\subsection{Comparison with previous analysis of the CMASS data with the MFs}
We note the previous analysis of the CMASS sample using the MFs reported a constraint of $\omega_{\rm cdm}=0.114\pm0.005$ after applying a Planck prior of  $n_s=0.965\pm0.004$ \cite{2022ApJ...928..108A}. A similar analysis of the two-dimensional genus extracted from shells of the BOSS data in \cite{Appleby_2020} found $\omega_{\rm cdm}=0.121\pm0.006$, also with a Planck prior on $n_s$. These constraints are derived from the amplitudes of the MFs, and only Gaussian information in the MFs is exploited. Our constraint on $\omega_{\rm cdm}$ is consistent with their measurements, and the extra non-Gaussian information improves the sensitivity on $\omega_{\rm cdm}$ by at least two times. At the same time, our analysis has included the degeneracies between $\omega_{\rm cdm}$ and $n_s$, and degeneracies with other cosmological parameters and HOD parameters. Our analysis also has a more rigorous treatment of the redshift-space distortions, Alcock-Paczynski distortions, and galaxy biasing. The tighter constraint we obtain highlights the importance of methodologies used in this analysis because they make it possible to fully exploit the wealth of valuable information captured by the MFs beyond the Gaussian part.

\subsection{Comparison with the Fisher matrix forecast.}
In \cite{2023JCAP...09..037L}, we performed a Fisher matrix analysis and quantified the constraining power of the MFs using the Molino mock galaxy catalogs \cite{2021JCAP...04..029H}, which are constructed from the halo catalogs of the Quijote simulations \cite{2020ApJS..250....2V} based on the HOD framework with parameters for the SDSS $M_r<-21.5$ and -22 galaxy samples. However, there still exists a large difference in the analysis between the Fisher forecast and this work. First, the galaxy number density of the Molino mocks is $\bar{n}\sim1.63\times 10^{-4}h^3\rm Mpc^3$, which is smaller than that of the galaxy samples used in this work ($\bar{n}=2.4\times 10^{-4}h^3\rm Mpc^3$). Therefore, the smallest smoothing scale used in the Fisher forecast ($20h^{-1}\rm Mpc$) is larger than that ($15h^{-1}\rm Mpc$) used here. In addition, the two analyses also differ in the cosmological and HOD parameter space, for example, the neutrino mass sum is considered in the Fisher analysis but not in this work. Finally, the power spectrum monopole and quadrupole are used to benchmark the constraining power of the MFs instead of the multipole moments of 2PCF used here. Despite these differences and focusing on the three cosmological parameters $h$, $n_s$, and $\sigma_8$, whose constraints are derived in both analyses. The Fisher forecast reported that the MFs provide 1.9, 2.0, and 1.3 times tighter constraints on $h$, $n_s$, and $\sigma_8$, respectively, compared to the power spectrum. This is comparable to our finding in this work: a factor of 2.0, 1.4, and 1.7 improvements come from the MFs on the three parameters relative to the 2PCF.

\section{Conclusions}
\label{sec:conclusion}
In this work, we present a simulation-based analysis of the Minkowski functionals for the CMASS sample from BOSS DR12, using an emulator for the MFs trained with forward-modeled galaxy mocks. We start from large and high-accuracy halo catalogs of the \textsc{Abacus}\textsc{Summit} simulation suite and construct high-fidelity mock galaxy catalogs using the Halo Occupation Distribution (HOD) framework, which includes 85 different cosmologies spanning an 8D parameter space and a total of 43521 variations in the 7D HOD parameter space. The effects of redshift-space distortions and Alcock–Paczynski distortions are applied during the forward modeling process, where we add layers of realism, including the complicated survey geometry and veto masks, mask the angular areas with low completeness, and eliminate the radial variation of number density by downsampling galaxies. We train a neural network emulator with the measured MFs from these forward-modeled galaxy mocks to emulate the cosmological and HOD dependences of the MFs and capture associated systematics. The same pipeline is repeated for the multipole moments of the galaxy 2-point correlation function (2PCF). Hence, we can use the emulator for the 2PCF to cross-check the robustness of the pipeline and benchmark the constraining power of the MFs.

Before the application of the full analysis pipeline to the CMASS data, we first validate the forward model with ten internal Abacus mocks at each of the four test cosmologies, which span a broad range of cosmological parameters. To further check the robustness and flexibility of our methodology, we also test the emulator against the Uchuu galaxy mock, which uses a SHAM framework to populate galaxies in halos from the Uchuu simulations. We succeed in recovering the true cosmological parameters from this mock, where the halo-galaxy connection model, halo finder, and the N-body code used are all different from those used in our forward model.

Applying our forward model to analyze the CMASS sample in the redshift range $0.45<z<0.58$,  we find the MFs outperform the galaxy 2PCF in the constraining power on $\omega_{\rm cdm}$, $\sigma_8$, and $n_s$ by factors of 2.0, 1.7, and 1.4, respectively. Combination of the MFs and 2PCF can further produce a $12\%$ and $18\%$ improvement in the constraint on $\sigma_8$ and $n_s$ from the MFs, but the 2PCF does not provide supplementary information on $\omega_{\rm cdm}$. Focusing on the combination of the MFs and 2PCF, our base $\Lambda$CDM analysis finds $\omega_{\rm cdm}=0.1172^{+0.0020}_{-0.0023}$, $\sigma_8=0.783\pm 0.026$, and $n_s=0.966^{+0.019}_{-0.015}$. The derived constraint $f\sigma_8=0.453 \pm 0.016$ is in agreement with Planck 2018 predictions and other results from a series of studies in the literature, and it is 1.9 times tighter than the result from the 2PCF.  The extended cosmological analysis reports $\alpha_s=-0.007^{+0.013}_{-0.017}$, $N_{\rm eff}=3.00^{+0.31}_{-0.37}$, $w_0=-0.98^{+0.07}_{-0.10}$, and $w_a=0.02^{+0.22}_{-0.20}$. Hence we don't find a significant departure from the $\Lambda$CDM model.

We note that a fixed prior on the acoustic scale $\theta_{*}$ from the Planck observations \cite{2020A&A...641A...6P} is implicitly adopted in our forward model because this constraint is used to set the Hubble parameter for the cosmological models used by the \textsc{Abacus}\textsc{Summit} simulations. This means we should interpret our constraint on $h$ and our comparison to other analyses with caution. However, the methodologies used in this analysis can be easily ported to other mocks constructed from cosmological simulations without this limitation. On the other hand, our emulators are based on the simulation snapshots at the redshift $z=0.5$, which is slightly different from the effective redshift of the CMASS sample used in this work, $z_{\rm eff} = 0.519$. The evolutions of galaxy clustering and halo-galaxy connection are neglected in our forward model. We plan to investigate these effects with the available Abacus lightcones \cite{2023MNRAS.525.4367H} in future work. In addition, the galaxy assembly bias is not implemented in our HOD model, but we find this effect will not significantly influence our results. We will study in detail how the assembly bias influences the MFs and its impact on cosmological parameter inference in future work.

There is still considerable room for improvement in the forward pipeline. First, neither the CMASS data nor the \textsc{Abacus}\textsc{Summit} simulation are fully utilized. For the CMASS sample, the redshift cut will first discard all galaxies in low-density regions. Then, the galaxies in high-density regions are downsampled, which also removes a large number of galaxies. The survey volume used is further reduced after the removal of regions near the boundary. For the simulation data, the cubic box is cut to match the survey geometry, which means a large fraction of the simulation volume is discarded. Although the volume usage is increased by repeatedly rotating and cutting the simulation box, the simulation data is still not fully used.  The development of a new pipeline is needed for more efficient usage of both observation and simulation data. On the other hand, we still rely on weights added to each galaxy to reduce the effect of fiber collisions. A detailed study and a more careful treatment of fiber collisions and other observational systematics are required for the exploration of information on smaller scales. 

The Minkowski functionals are popular alternative statistics for the analysis of LSS data since its introduction to cosmology. Recently, a series of our works have emphasized the promising prospects of using the MFs of LSS as a new probe of modified gravity \cite{2017PhRvL.118r1301F,2023arXiv230504520J} and massive neutrinos \cite{2022JCAP...07..045L,2023JCAP...09..037L} (also see Liu et al. \cite{PhysRevD.101.063515}). Although the constraining power of the MFs on cosmological parameters in the concordance $\Lambda$CDM model and its extensions is promising, there are no available theoretical predictions for the MFs that are accurate enough on small scales. This work lays the foundation for the application of the MFs on small scales to spectroscopic galaxy data and again demonstrates their strong constraining power on cosmological parameters. Current and upcoming galaxy redshift surveys such as DESI \cite{desicollaboration2016desiexperimentisciencetargeting}, the Nancy Grace Roman Space Telescope \cite{green2012widefieldinfraredsurveytelescope}, Euclid \cite{laureijs2011eucliddefinitionstudyreport} and CSST \cite{CSST,2019ApJ...883..203G} will provide high-precision measurements for galaxies.  The vast amount of information extracted with higher-order statistics like the MFs will tighten cosmological parameter constraints and improve our understanding of both galaxy formation and cosmology.

\acknowledgments
We would like to express our heartfelt gratitude to Jeremy L. Tinker for generously generating the Uchuu mock for us. We are grateful to Zhongxu Zhai for kindly helping us transfer the Uchuu mock data. WL also thanks Zhongxu Zhai, Cheng Zhao, and Aoxiang Jiang for helpful discussions. We are also grateful to the anonymous referee for the constructive comments, which helped to improve the quality of this paper. This work is supported by the National Natural Science Foundation of China Grants No. 12173036 and 11773024, by the National Key R\&D Program of China Grant No. 2021YFC2203100 and No. 2022YFF0503404, by the China Manned Space Program with Grant No. CMS-CSST-2025-A04 and Grant No. CMS-CSST-2021-B01, by the Fundamental Research Funds for Central Universities Grants No. WK3440000004 and WK3440000005, by Cyrus Chun Ying Tang Foundations, and by the 111 Project for "Observational and Theoretical Research on Dark Matter and Dark Energy" (B23042). EP was in part supported by the U.S. Department of Energy HEP-AI program grant DE-SC0023892.

\appendix

\section{Gaussianity test}
\label{sec:Gaussian_test}
\begin{figure}[tbp]
	\centering
	\includegraphics[width=1.0\textwidth]{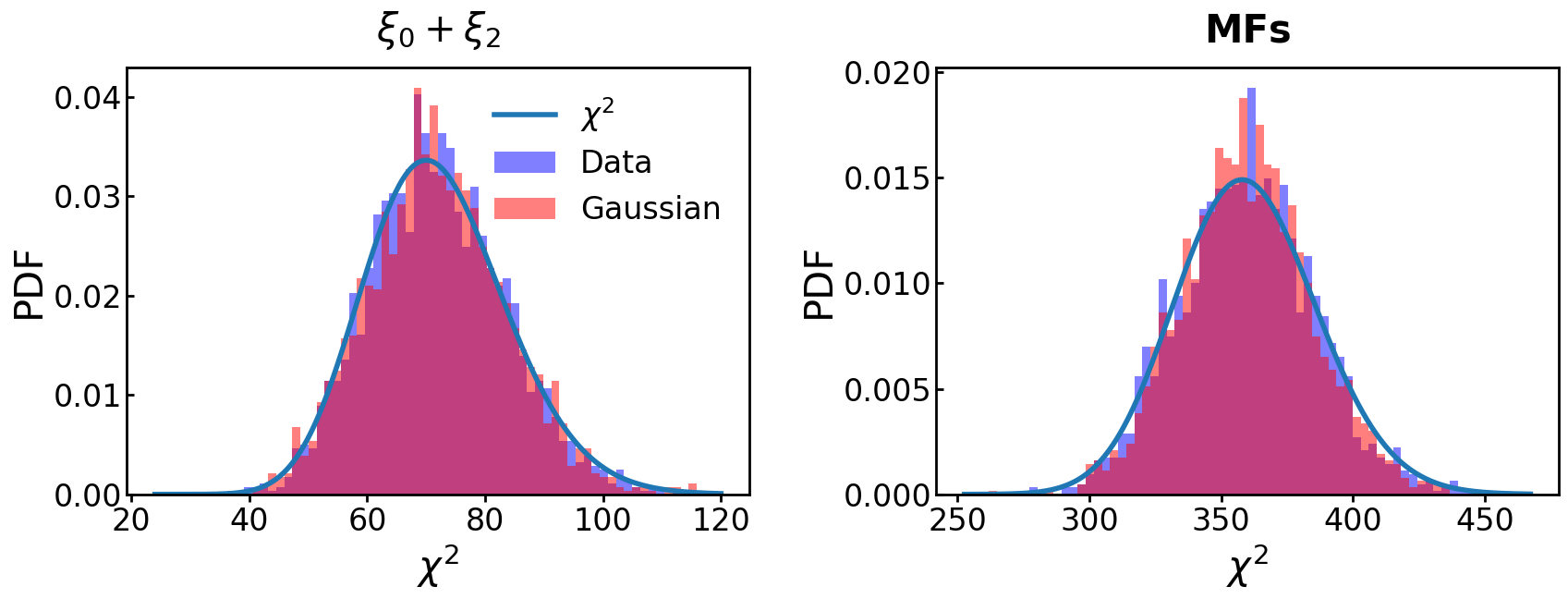}
	\caption{\label{fig:gaussian_test} A qualitative assessment of the Gaussianity of the likelihood for the monopole and quadrupole moments of the 2-point correlation function ($\xi_{0}+\xi_2$, left column) as well as the Minkowski functionals (MFs, right column). The histograms of the $\chi^2$ values measured from the 2048 Patchy mocks are drawn in blue, while those of the $\chi^2$ values measured from a multivariate Gaussian distribution with the same mean and covariance as the Patchy mocks are shown in red. The solid blue lines show theoretical $\chi^2$ distributions with degrees of freedom equal to the total number of observables (72 for $\xi_{0}+\xi_2$ and 360 for MFs).} 
\end{figure}

We assume a Gaussian likelihood for both the MFs and 2PCF in the analysis presented in this work. To validate this assumption, we follow the analysis performed in \cite{10.1093/mnras/stab2384,Paillas_2023} and check that the likelihood of these statistics can be approximated by the multivariate Gaussian. We perform this Gaussianity test with the 2048 Patchy mocks. Hence, we can obtain 2048 $\chi^2$ values for each of the statistics by 
\begin{equation}
\chi^2_i=(\boldsymbol{x}_{\rm patchy}^i-\overline{\boldsymbol{x}}_{\rm patchy})^T C^{-1} (\boldsymbol{x}_{\rm patchy}^i-\overline{\boldsymbol{x}}_{\rm patchy}),
\end{equation}
where $\boldsymbol{x}_{\rm patchy}^i$ is the data vector of the summary statistics for the $i$-th Patchy mock catalog, $\overline{\boldsymbol{x}}_{\rm patchy}$ and $C$ is the mean and the covariance matrix of the data vector estimated from the Patchy mocks. 

If the assumption of Gaussian likelihood holds, the $\chi^2$ values are expected to follow a $\chi^2$ distribution with degrees of freedom equal to the length of the data vector. In figure~\ref{fig:gaussian_test}, we plot the histogram (in blue) of the $\chi^2$ values measured from the Patchy mocks and compare it with the theoretical $\chi^2$ distribution curve for each of the summary statistics. Due to the existence of fluctuations in the histogram, the curve may not exactly agree with the histogram, even for a sample strictly following the $\chi^2$ distribution. To visualize the amplitude of the fluctuations around the theoretical $\chi^2$ distribution curve and use it as a ruler to help assess the Gaussianity of the data vectors, we create 2048 multivariate Gaussian distributed data vectors with the same mean and covariance matrix as those estimated from the Patchy mocks. We then obtain 2048 $\chi^2$ values for the multivariate Gaussian distributed data vectors and also plot a histogram (in red) for them in figure~\ref{fig:gaussian_test}. As seen in this figure, the histogram of $\chi^2$ values for all statistics is very close to that for the Gaussian distributed data vectors and agrees well with the theoretical $\chi^2$ distribution curve. This indicates that the likelihood of the 2PCF and MFs can be well modeled as Gaussian.

\section{Test of pipeline}
\label{sec:pipetest}
\begin{figure}[tbp]
	\centering
	\includegraphics[width=1.0\textwidth]{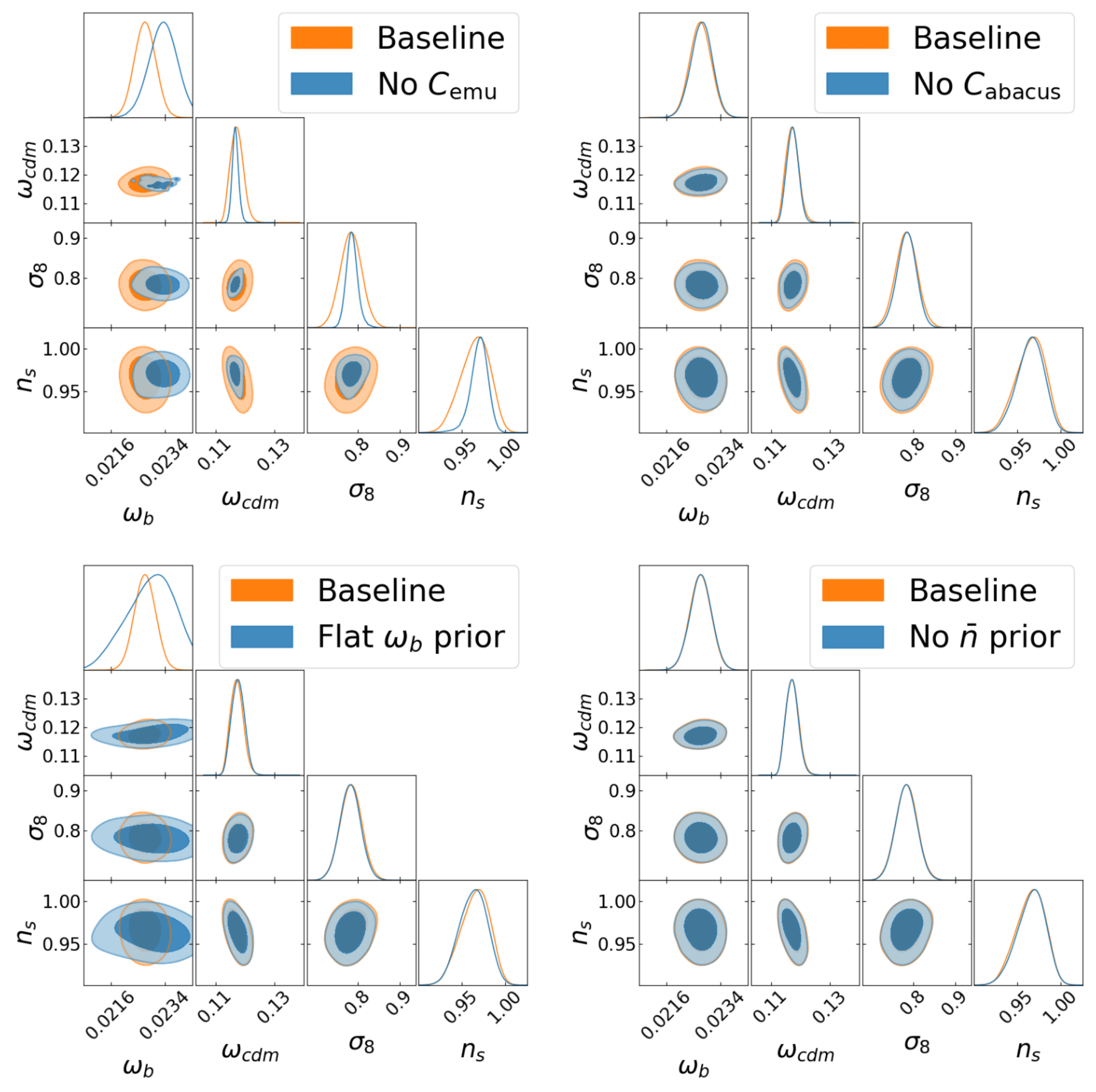}
	\caption{\label{fig:Fit_setting_test} Comparison of marginalized constraints on $\omega_{b}$, $\omega_{\rm cdm}$, $\sigma_8$, and $n_s$ obtained from the baseline likelihood analysis with those from analyses where the emulator error is not included (top left panel, ``No $C_{\rm emu}$''), the sample variance of the galaxy mocks used to train the emulator is neglected (top right panel, ``No $C_{\rm abacus}$''), the Gaussian BBN prior on $\omega_{\rm b}$ is not used (bottom left panel, ``Flat $\omega_{\rm b}$ prior''), and the prior imposed by the mean number density constraint is not adopted (bottom right panel, ``No $\bar{n}$ prior''). All of the posterior distributions are calculated from the combination of the MFs and 2PCF.} 
\end{figure}

In this appendix, we examine how the variations in our baseline likelihood analysis impact the posterior probability distribution from the combination of the MFs and 2PCF. We compare the marginalized posterior distribution obtained with the baseline analysis and four different cases in figure~\ref{fig:Fit_setting_test}: 1, without the emulator error; 2, without the sample variance from the Abacus mocks; 
3, without the BBN prior on $\omega_{\rm b}$; 4, without the explicit prior imposed by the mean number density constraint.

As explained in Section~\ref{sec:likelihood}, the total covariance matrix collects contributions from the sample variance of the CMASS data $C_{\rm data}$, the emulator error $C_{\rm emu}$, and the sample variance of the mock catalogs used to train the emulator $C_{\rm abacus}$. The relative contributions from each of the three sources are visualized in figure~\ref{fig:C_frac}. From the first row of figure~\ref{fig:Fit_setting_test}, we find the exclusion of the emulator error $C_{\rm emu}$ can lead to a significant change in the marginalized posteriors for the four cosmological parameters $\omega_{b}$, $\omega_{\rm cdm}$, $\sigma_8$, and $n_s$. The constraints on $\omega_{\rm cdm}$, $\sigma_8$, and $n_s$ are much tighter when neglecting the emulator error, which indicates considerable room for improvement in the training of the emulators and the constraining power of these summary statistics. On the other hand, the shift of the inferred $\omega_{\rm b}$ to a higher value than the input BBN prior demonstrates the necessity of the inclusion of the $C_{\rm emu}$ term in the total covariance matrix. Whether the Abacus sample variance $C_{\rm abacus}$ is added into the error budget or not does not make a big difference in the obtained posterior distributions. This is understandable, considering the relatively small importance of this term seen in figure~\ref{fig:C_frac}.

Our baseline analysis used a tight BBN prior on $\omega_{\rm b}$. From the bottom left panel of figure~\ref{fig:Fit_setting_test}, we find the constraint on $\omega_{\rm b}$ is significantly weakened without this prior, but the constraints on the other three cosmological parameters are almost not impacted. As seen from the bottom right panel, the explicit prior imposed by the mean number density constraint defined in Eq~\ref{eq:barn_prior} does not have a perceivable influence on the constraints of the four cosmological parameters as well. This is because our target number density for the Abacus mocks is very low. Hence, this prior is not strong enough to make a difference. However, we find its influence is larger in the constraints on the HOD parameters, as expected. For the analysis of a denser galaxy sample, we find this prior will be more important for the summary statistics like the MF, which are sensitive to the number density of galaxies.

\section{Constraints of HOD parameters}
\label{sec:HOD_constraints}

\begin{center}
	\small
	\begin{table}[tbp]
	\scalebox{1.1}[1]{
    	\begin{tabular}{ccccccc}
            \hline & \multicolumn{2}{c}{ MFs } & \multicolumn{2}{c}{ 2PCF } & \multicolumn{2}{c}{ 2PCF + MFs } \\
            \hline Parameter & best-fit & mean $\pm \sigma$ & best-fit & mean $\pm \sigma$ & best-fit & mean $\pm \sigma$ \\
            \\[-1em]
            $logM_{cut}$ & 12.77 & $12.78^{+0.05}_{-0.06} $ & 13.07& $12.80\pm 0.11    $ & 12.77& $12.78^{+0.05}_{-0.06}  $\\
            \\[-1em]
            $logM_1$ & 13.72 & $13.61\pm 0.20 $ &13.52& $14.03^{+0.31}_{-0.39}             $ & 13.79 & $13.60\pm 0.19     $\\
            \\[-1em]
            $log\sigma$ & -2.87& $-1.71^{+0.90}_{-1.1}$ &-0.50& $-1.6^{+1.1}_{-1.3}$ & -2.90&$-1.75^{+0.91}_{-1.1}$\\
            \\[-1em]
            $\alpha$ & 1.05& $0.85^{+0.15}_{-0.28} $ &0.58& $0.92^{+0.16}_{-0.40}$ &1.05& $0.84^{+0.14}_{-0.28}     $\\
            \\[-1em]
            $\kappa$ &3.3& $4.9^{+1.5}_{-1.3}$ &6.9& $4.3^{+3.1}_{-1.7} $ &2.7& $5.0^{+1.6}_{-1.3}       $\\
            \\[-1em]
            $\alpha_{vel,c}$ &0.24& $<0.46$ &0.16& $0.28^{+0.14}_{-0.12}   $ & 0.21& $0.19\pm 0.10    $\\
            \\[-1em]
            $\alpha_{vel,s}$ & 0.95&$0.77^{+0.47}_{-0.34} $ &1.39& $>1.10$ &1.09& $1.02\pm 0.16      $\\
            \\[-1em]
            \hline
        \end{tabular}
        }
    \caption{\label{tab:hod_constraints} The best-fit, mean, and $68\%$ confidence interval values of the seven HOD parameters for the MFs, 2PCF, and their joint combination, marginalized over cosmological parameters.}
    \end{table}
\end{center}

\begin{figure}[tbp]
	\centering
	\includegraphics[width=1.0\textwidth]{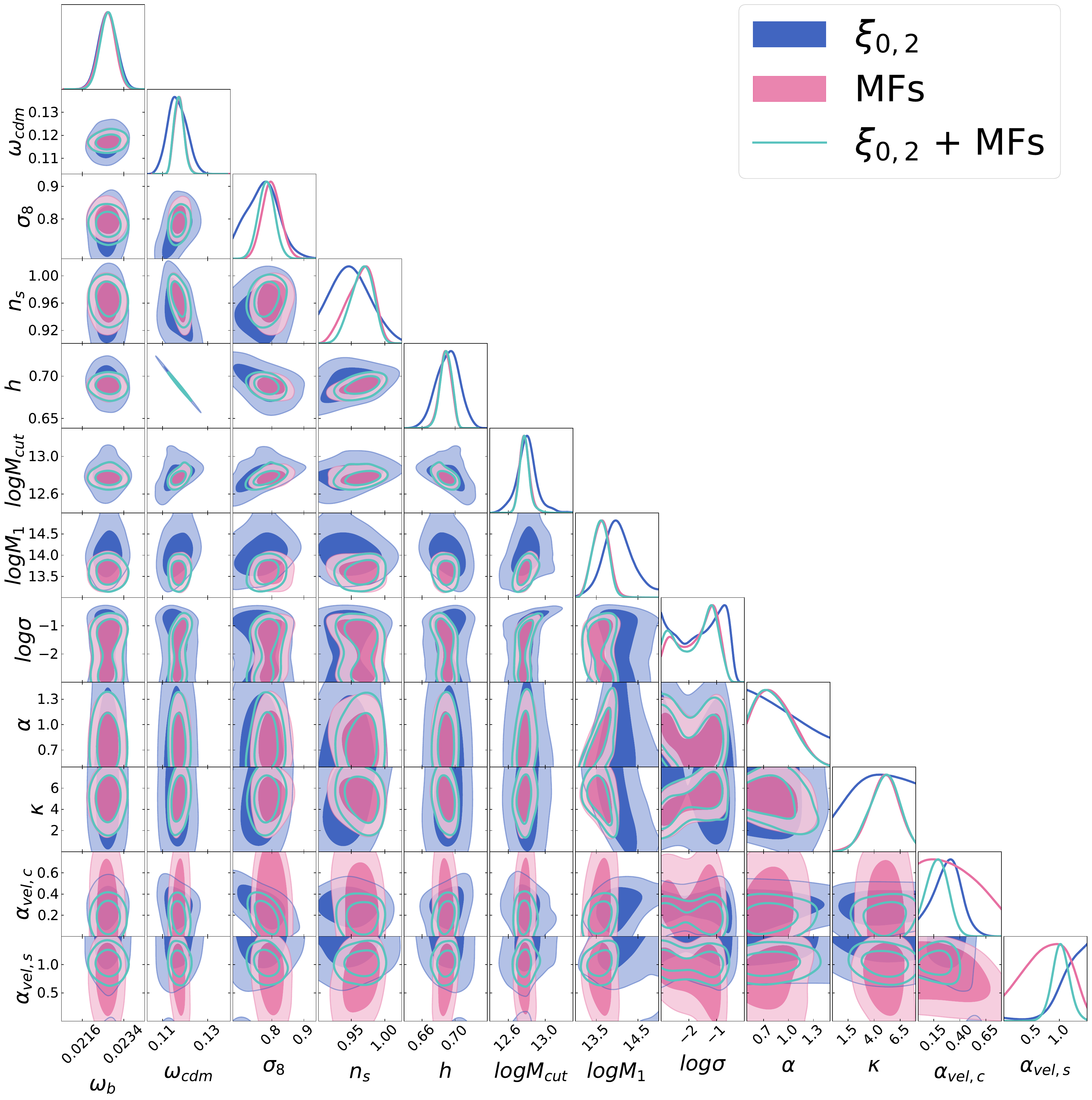}
	\caption{\label{fig:CMASS_fit_cos_hod} Marginalized constraints on the five cosmological and seven HOD parameters from the 2PCF, the MFs, and the combination of the 2PCF and MFs for the BOSS CMASS galaxy sample. Also, a BBN Gaussian prior $\mathcal{N}\left(0.02268, 0.00038\right)$ is used for $\omega_{\rm b}$.} 
\end{figure}

We plot the posterior distribution of the full cosmological and HOD parameters from the 2PCF, the MFs, and the combination of the 2PCF and MFs for the BOSS CMASS galaxy sample in figure~\ref{fig:CMASS_fit_cos_hod}. The corresponding best-fit, mean, and $68\%$ confidence interval values of the HOD parameters are also listed in table~\ref{tab:hod_constraints}. We find the MFs are more sensitive to all of the HOD parameters except for the velocity bias parameters $\alpha_{vel,c}$ and $\alpha_{vel,s}$. There are two reasons for this. On one hand, a considerable proportion of the velocity information related to the velocity bias parameters is smeared out by the isotropic Gaussian smoothing process since we have used a relatively large smoothing scale $R_G=15~h^{-1}\rm Mpc$. Therefore, if the change of velocity bias parameter only leads to variations in the LSS with scales much smaller than the smoothing scale, the MFs can fail to detect its impact and the sensitivity of the MFs to this change is low. We find the constraining power of the MFs on the two velocity bias parameters can be increased when small smoothing scales are used. On the other hand, the MFs collect contributions from the infinitesimal volume or area element in an angle-independent way, as seen from equation~\ref{eq:mfs_def}, this means the anisotropic information can be buried when adding up contributions that involve the LoS direction and those that involve directions orthogonal to the LoS \cite{2013MNRAS.435..531C,2021arXiv210803851J}. Therefore, when the lost anisotropic information is added by the 2PCF (in particular, $\xi_2$), we find the constraints from the MFs on $\alpha_{vel,c}$ and $\alpha_{vel,c}$ are improved.

We find our constraints on $logM_{cut}$, $logM_1$, $log\sigma$, and $\alpha$ are in agreement with the baseline results of \cite{10.1093/mnras/stab3355}. However, their measurement for $\kappa$ is much lower than our inference. The joint analysis of the two statistics infers $\alpha_{vel,c}=0.19\pm0.10$ and $\alpha_{vel,s}=1.02\pm0.16$, that is, a nonzero central velocity bias with $\sim 2\sigma$ significance and a satellite velocity bias consistent with 1. This measurement is consistent with the results of \cite{10.1093/mnras/stab3355,10.1093/mnras/stu2120}, both of the two works found $\alpha_{vel,c} \simeq 0.2$ and $\alpha_{vel,c} \simeq 1.0$. We note that the galaxy assembly bias is not implemented in the analysis, which will be investigated in our future work. The mass resolution of the fiducial cosmology in the \textsc{Abacus}\textsc{Summit} simulation \cite{2021MNRAS.508.4017M} is $2 \times 10^9 h^{-1} \mathrm{M}{\odot}$. Since the mass resolution is proportional to $\Omega_m$, and the varying range of $\Omega_m$ is approximately $\pm 20\%$, the simulations for non-fiducial cosmologies maintain a similar mass resolution. The best-fit values for $log M_{cut}$ and $log M_{1}$, obtained from the combination of the 2PCF and MFs, are 12.77 and 13.79, respectively. Therefore, the mass resolution of the \textsc{Abacus}\textsc{Summit} simulation is more than sufficient to resolve the lightest halos hosting central and satellite galaxies for the CMASS galaxy sample.

\begin{figure}[tbp]
	\centering
	\includegraphics[width=1.0\textwidth]{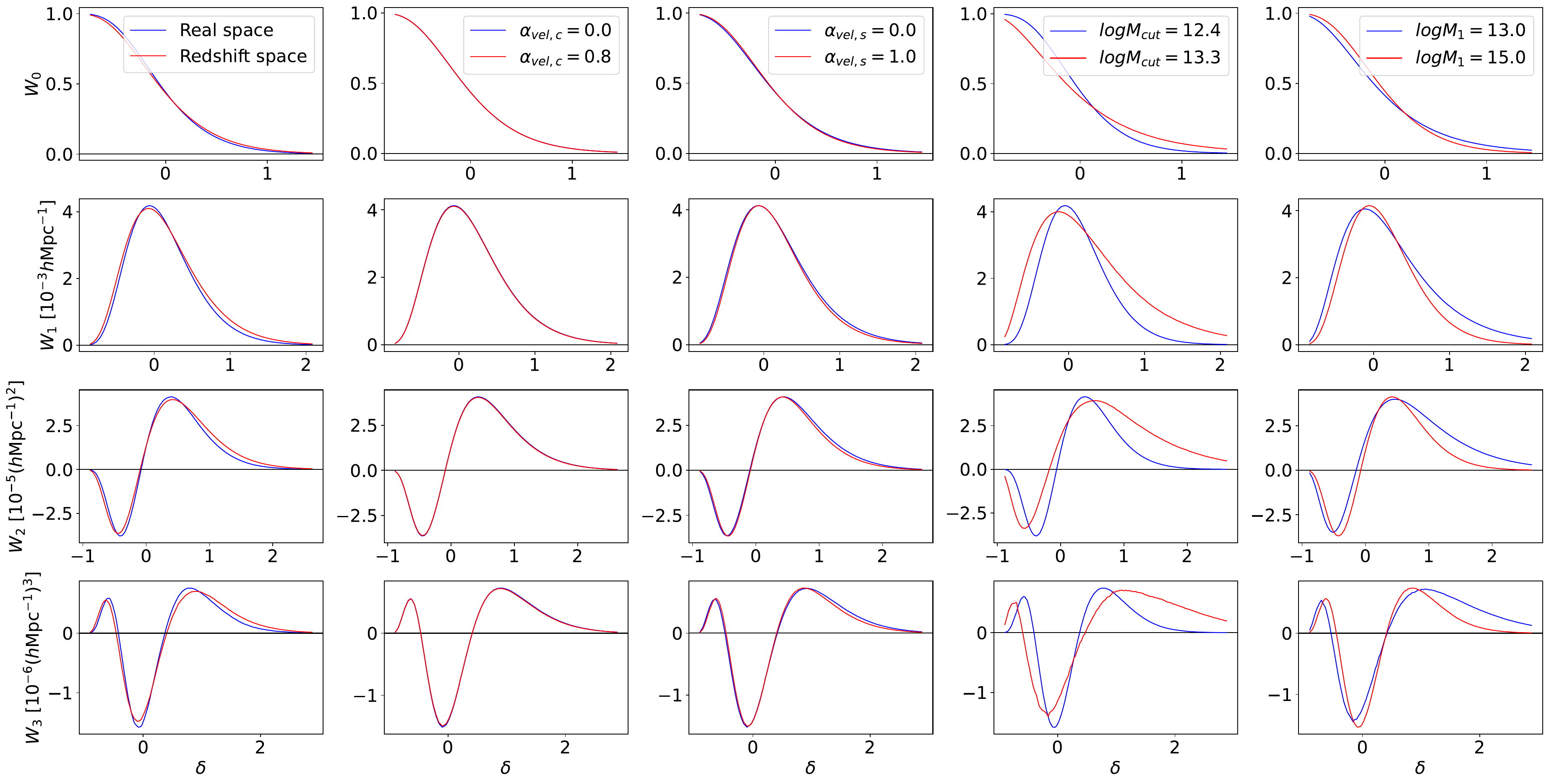}
	\caption{\label{fig:GalaxyBias_vs_VelocityBias} The MFs of the HOD mock in real and redshift space (first column), with and without central (second column) or satellite velocity bias (third column), with different values of $log M_{cut}$ (fourth column) or $logM_1$ (last column). The values of other HOD parameters are set to the best value from the MFs.} 
\end{figure}

To study the effect of $\alpha_{cel,s}$ and $\alpha_{cel,c}$ on the MFs and disentangle their effects from other HOD parameters such as $logM_{cut}$ and $logM_1$, we plot in figure~\ref{fig:GalaxyBias_vs_VelocityBias} the MFs of the HOD mocks in both real and redshift space, with and without central or satellite velocity bias, and for different values of $log M_{cut}$ or $logM_1$. We compare the difference in the MFs between real space and redshift space here because the effect of velocity bias is related to redshift space distortions (see \cite{2021arXiv210803851J} for a detailed interpretation of the effect of RSD on the MFs). We find the effect of RSD on the MFs is reduced when $\alpha_{vel,c}=0.8$ \footnote{This phenomenon becomes more pronounced when we increase $\alpha_{vel,c}$ further. However, in this case, we aim to show the differences in the MFs caused by parameter variations within the hard prior range used in this analysis.} or $\alpha_{vel,s}=1.0$. In these cases, the relative motions of both central and satellite galaxies with respect to their host halos cause the density field in redshift space to resemble that in real space to some extent. This finding is consistent with \cite{10.1093/mnras/stu2120}, where Fig. 6 shows that a low satellite velocity bias tends to make the contours more squashed along the line-of-sight direction, while a high central velocity bias has the opposite effect. Additionally, we observe a (weak) anticorrelation between $\alpha_{vel,c}$ and $\alpha_{vel,s}$ in the bottom right corner of our Figure 15, similar to what is seen in their Fig. 5. This anticorrelation arises because $\alpha_{vel,c}$ and $\alpha_{vel,s}$ can partially compensate for each other.
 
On the other hand, $log M_{cut}$ and $log M_1$ affect the MFs in a different manner. The most notable feature is that the MFs curves either expand or compress as the values of these two parameters change. Since $log M_{cut}$ controls the halo mass required to host a central galaxy, a larger $log M_{cut}$ corresponds to a higher linear galaxy bias. As a result, we observe more extended MFs curves for $log M_{cut} = 13.3$ compared to $log M_{cut} = 12.4$. The effect of $log M_1$ is more intricate: although smaller $log M_1$ allows halos with lower masses to host satellite galaxies, larger halos will also accommodate more satellite galaxies (since the power-law index for the number of satellite galaxies, $\alpha$, is larger than one for the CMASS galaxies). This means that a larger fraction of galaxies can be hosted by heavier halos in the HOD mock with smaller $log M_1$. Consequently, it is understandable that the MFs curves exhibit a higher galaxy bias for $log M_1 = 13.0$ than for $log M_1 = 15.0$. $log M_{cut}$ and $log M_1$ can also affect the amplitude of the MFs, as they influence the shape of the power spectrum (see Section 2 of \cite{2014MNRAS.441L..21H} for an analytical description of how these parameters impact the power spectrum).

We hope that the explanation of how the four HOD parameters influence the MFs in different ways will help readers understand how the MFs can infer the velocity bias parameters, along with the other HOD parameters that determine galaxy bias.

\bibliographystyle{JHEP}
\bibliography{reference}
\end{document}